\documentclass{article}

\usepackage{PRIMEarxiv}

\usepackage[utf8]{inputenc} 
\usepackage[T1]{fontenc}    
\usepackage{hyperref}       
\usepackage{url}            
\usepackage{booktabs}       
\usepackage{amsfonts}       
\usepackage{nicefrac}       
\usepackage{microtype}      
\usepackage{lipsum}
\usepackage{fancyhdr}       
\usepackage{graphicx}       
\graphicspath{{media/}}     
\usepackage[title]{appendix}

\usepackage[ruled,linesnumbered, noend]{algorithm2e} 
\SetKwInput{KwInput}{Input}     
\SetKwInput{KwOutput}{Output}

\title{\MakeLowercase{t}-METASET: Tailoring Property Bias of
Large-Scale Metamaterial Datasets through
Active Learning}

\author{
  Doksoo Lee \\
  Dept. of Mechanical Engineering \\
  Northwestern University \\
  Evanston, IL 60208\\
  \texttt{dslee@northwestern.edu} \\
   \And
  Yu-Chin Chan \\
  Siemens Corporate Technology \\
  Princeton, New Jersey 08540 \\
  \texttt{yu-chin.chan@siemens.com} \\
    \And
  Wei (Wayne) Chen \\
  Dept. of Mechanical Engineering \\
  Northwestern University \\
  Evanston, IL 60208 \\
  \texttt{wei.wayne.chen@northwestern.edu} \\    
    \And
 Liwei Wang \\
  School of Mechanical Engineering \\
  Shanghai Jiao Tong University \\
  Shanghai, P.R. China, 200240 \\
  \texttt{iridescence@sjtu.edu.cn} \\   
     \And
  Anton van Beek \\
  School of Mechanical and Materials Engineering \\
  University College Dublin \\
  Belfield, Dublin 4, Ireland, D04 V1W8\\
  \texttt{anton.vanbeek@ucd.ie} \\
    \And
  Wei Chen\thanks{Corresponding author: Wei Chen (weichen@northwestern.edu)} \\
  Dept. of Mechanical Engineering \\
  Northwestern University \\
  Evanston, IL 60208\\
  \texttt{weichen@northwestern.edu} \\    
}

\usepackage{fancyhdr}
\usepackage{upgreek}
\usepackage{graphicx} 
\usepackage{hyperref}   
\usepackage{fancyhdr,amssymb} 
\usepackage{float}
\usepackage{xcolor}
\usepackage{soul}
\hypersetup{
	colorlinks=true,
	linkcolor=blue,
	citecolor=blue,
	urlcolor=blue,
}
\hypersetup{final} 
\usepackage[square,numbers]{natbib}

\usepackage{amsmath, bm} 
\usepackage{amsfonts} 
\usepackage{enumitem}
\usepackage{mathtools}
\usepackage{calrsfs}
\usepackage[utf8]{inputenc} 
\usepackage[T1]{fontenc}    

\DeclareMathAlphabet{\mathcal}{OMS}{cmsy}{m}{n}

\begin{document}
\maketitle

\begin{abstract}
Inspired by the recent achievements of machine learning in diverse domains, data-driven metamaterials design has emerged as a compelling paradigm that can unlock the potential of multiscale architectures. The model-centric research trend, however, lacks principled frameworks dedicated to data acquisition, whose quality propagates into the downstream tasks. Often built by naive space-filling design in shape descriptor space, metamaterial datasets suffer from property distributions that are either highly imbalanced or at odds with design tasks of interest. To this end, we present t-METASET: an active-learning-based data acquisition framework aiming to guide both diverse and task-aware data generation. Distinctly, we seek a solution to a commonplace yet frequently overlooked scenario at early stages of data-driven design of metamaterials: when a massive ($\sim O(10^4)$) shape-only library has been prepared with no properties evaluated. The key idea is to harness a data-driven shape descriptor learned from generative models, fit a sparse regressor as a start-up agent, and leverage metrics related to diversity to drive data acquisition to areas that help designers fulfill design goals. We validate the proposed framework in three deployment cases, which encompass general use, task-specific use, and tailorable use. Two large-scale mechanical metamaterial datasets are used to demonstrate the efficacy. Applicable to general image-based design representations, t-METASET could boost future advancements in data-driven design.
\end{abstract}



\section{Introduction}
Metamaterials are artificially architectured materials that support unusual properties from their structure rather than composition~\cite{yu2018mechanical}. The recent advancements of computing power and manufacturing have fueled research on metamaterials, including theoretical analysis, computational design, and experimental validation. Over the last two decades, outstanding properties and functionalities achievable by metamaterials have been reported from a variety of fields, such as optical~\cite{soukoulis2011past}, acoustic~\cite{cummer2016controlling}, thermal~\cite{schittny2013experiments}, and mechanical~\cite{kadic2013metamaterials}. They have been widely deployed to applications in communications, aerospace, biomedical, and defense, to name a few~\cite{liu2015metamaterials}. From a design point of view, leveraging the rich designability in hierarchical systems is key to further disseminating metamaterials as a versatile material platform, which not only realizes superior functionalities but also facilitates customization and miniaturization. There has been growing demand for advanced design methods to harness the potential of metamaterials. 

Data-driven metamaterials design (DDMD) offers a route to intelligently design metamaterials. In general, the approach builds on three main steps: data acquisition, model construction, and inference for design purposes. DDMD typically starts with a precomputed dataset that includes a large number of structure-property pairs~\cite{zhu2017two,liu2018generative, ma2019probabilistic, wang2020deep,chan2021metaset}. Machine learning model construction follows to learn the underlying mapping from structure to property, and sometimes vice versa. Then the data-driven model is used for design optimization, such as at the ``building block” or unit cell level, and optionally tiling in the macroscale as well when aperiodic designs are of interest~\cite{wang2022data, chan2022remixing, da2022data, wang2021mechanical}. The key distinctions of DDMD against conventional approaches are that (i) DDMD accelerates multiscale design optimization via exploring the vast design space efficiently; (ii) it has little restrictions on analytical formulations of design interest; and (iii) some DDMD approaches enable on-demand design without iterations, which pays off the initial cost of data acquisition and model construction. Capitalizing on the advantages, DDMD has reported a plethora of achievements for diverse design problems in recent years~\cite{yu2018mechanical,liu2018generative, ma2019probabilistic, wang2020deep, wang2019robust, so2019simultaneous, gurbuz2021generative}.

Despite the recent surge of DDMD, rare attention has been given to data acquisition and data quality assessment -- the very first step of DDMD. In data-driven design, \textit{data is a design element}; a collection of data points forms a landscape to be learned by a model, which is an ``abstraction” of the data, and to be explored by either model inference or modern optimization methods. Hence data quality ends up propagating into the subsequent stages. Yet the downstream impact of naive data acquisition is opaque to diagnose and thus challenging to prevent \textit{a priori}~\cite{sambasivan2021everyone}. Underestimating the risk, common practice in DDMD typically resorts to a large number of space-filling designs in the shape space spanned by the shape parameters. This \textit{inevitably} hosts imbalance -- distributional bias of data -- in the property space~\cite{wang2022data, wang2021ih, chan2021metaset, wang2021data} formed by the property components. The downstream tasks involving a data-driven model -- training, validation, and deployment to design -- follow mostly without rigorous assessment on data quality in terms of diversity, design quality, and feasibility, among others. The practice overlooks not only \textit{data imbalance itself} but also \textit{the compounding ramification} at the design stage, allowing both to impede solid deployment of DDMD.

To this end, Chan et al. presented METASET~\cite{chan2021metaset} as a subset selection framework that can identify small yet diverse subsets from a fully evaluated database. 
Key idea is to evaluate the properties of \textit{all} the designs \textit{a priori}, then downsample a balanced subset based on diversity metrics. Yet the approach lacks generality of data acquisition for DDMD in that: (i) design evaluation could be prohibitively expensive to build a massive ($\sim O(10^4)$) database with all the data evaluated; (ii) diversity alone does not offer data customization for specific design tasks.

To enhance the generality and efficiency of data acquisition for DDMD, we propose \textit{task-aware} METASET (t-METASET) with special attention to starting with sparse observations. Herein, ``task-aware” approaches rate individual data points based on the utility for a given \textit{specific} design scenario, rather than on distributional metrics (e.g., diversity) for \textit{general} use. The proposed framework handles data bias reduction (for generic use) and design quality (for particular use) simultaneously, by leveraging diversity and quality as the sampling criteria, respectively. We advocate that (i) building a good dataset should be an \textit{iterative} procedure~\cite{ng2021chat, strickland2022andrew}; (ii) diversity sampling~\cite{kulesza2012determinantal} can efficiently suppress the property bias of multi-dimensional regression involved in most DDMD methods~\cite{chan2021metaset}; (iii) property bias control significantly improves fully aperiodic metamaterial designs, as shown by recent reports~\cite{wang2022data, chan2022remixing, da2022data, wang2021mechanical}. Distinct from existing work, however, we primarily seek a solution to a commonplace -- yet frequently overlooked -- scenario that designers face during data preparation: a large-scale shape dataset has been generated and is about to be observed \textit{without evaluated observations at the beginning}. 

\begin{figure*}[t]
\centering
\includegraphics[width=0.8\linewidth]{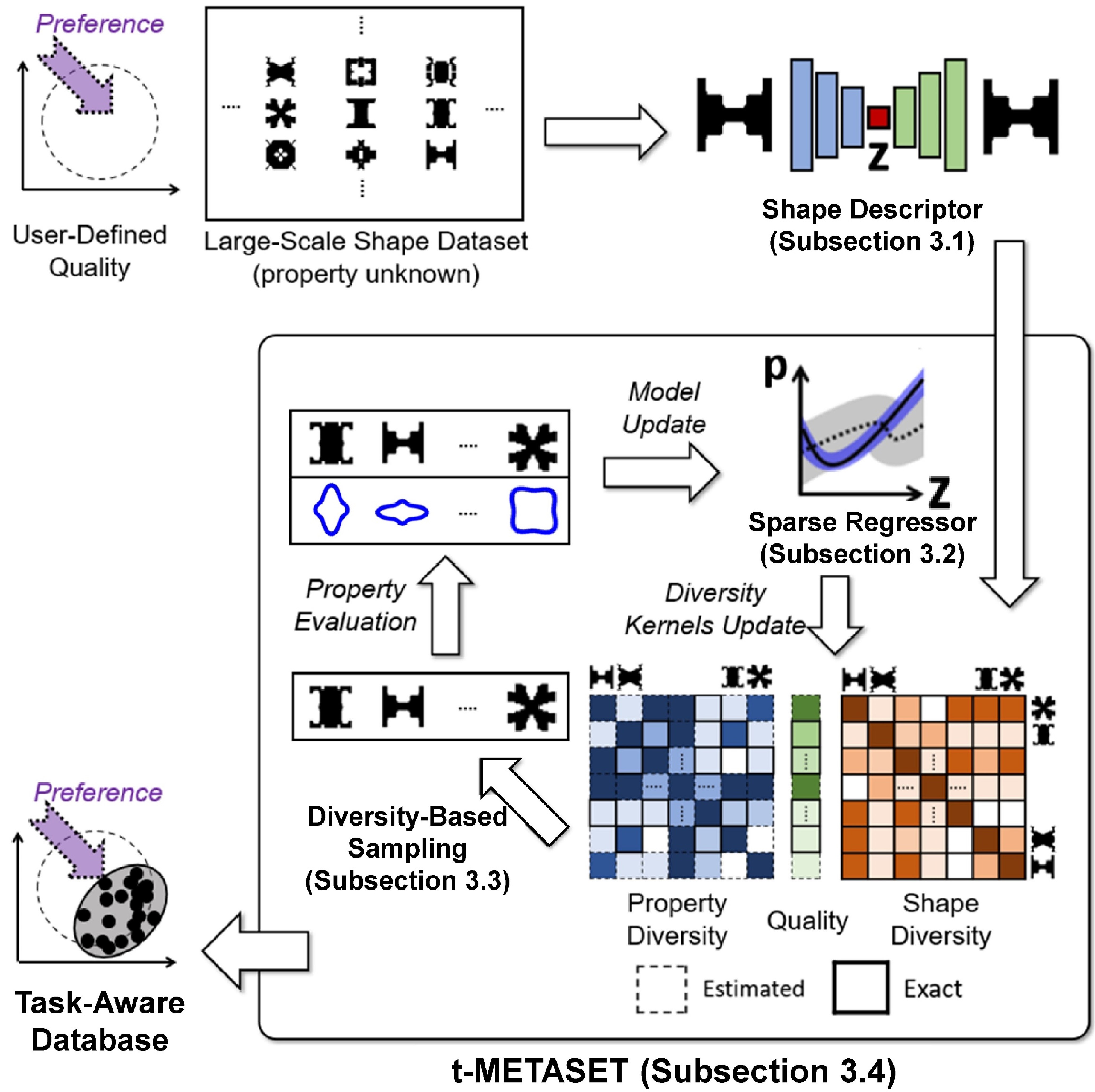}
\caption{An overview of t-METASET. Given a shape-only dataset, a compact shape descriptor of microstructures is distilled by a generative model (e.g., variational autoencoder). A sparse regressor (e.g., a Gaussian process) learns the descriptor-property mapping in light of sparse observations. Harnessing the prediction over unseen shapes, diversity sampling (e.g., Determinantal Point Processes) drives active learning through diversity and, optionally, task-related quality metrics. Once evaluated, the observed batch refines the regressor. By repeating this routine, t-METASET suppresses undesirable distributional bias while boosting desirable one.}
\label{fig:t-METASET}
\end{figure*}

Our t-METASET incrementally ``grows” a high-quality dataset that is not only diverse but also task-aware. Figure~\ref{fig:t-METASET} illustrates a schematic of the t-METASET procedure. The central ideas are (i)~to extract a compact shape descriptor from a shape-only dataset by unsupervised representation learning, (ii)~to sequentially update a sparse regressor as a start-up ``agent” under \textit{sparse} observations, and (iii)~to intelligently curate samples based on the prediction of the regressor, and batch sequential sampling~\cite{kulesza2012determinantal} building on shape diversity, \textit{estimated} property diversity, and user-defined quality. Starting from a massive library of building blocks, the active learning framework maneuvers the data acquisition so that it can tailor the data distribution based on both diversity (for generic use) and quality (for specific use) for given tasks.

In the context of DDMD, the intellectual contributions of t-METASET are three-fold:
\begin{itemize}[label=\textbullet]
    \item \textit{Starting without evaluated designs}, t-METASET offers a principled framework on how to build a diverse dataset \textit{during} data acquisition with rigorous metrics and a small amount of heuristics; 
    \item The framework provides a solution to \textit{property bias} that both existing and newly created metamaterial datasets are prone to; 
    
    \item The proposed t-METASET can produce \textit{task-aware} datasets whose distributional characteristics can be tailored in response to user-defined design tasks, while securing shape and property diversity along the way.
\end{itemize}

We argue the advantages of t-METASET are: (i) scalability, (ii) modularity, (iii) customizability to general or specific tasks, (iv) freedom from restrictions on shape generation schemes, (v) no dependency on domain knowledge and, by extension, (vi) applicability over generic design datasets involving high-dimensional images. t-METASET is validated via two large-scale shape-only mechanical metamaterial datasets (containing 88,180 and 57,000 instances, respectively) that are built from different ideas, without preliminary downsampling. The validation involves three scenarios addressed by different sampling criteria: (i) only diversity aiming at general use (e.g., global metamodeling~\cite{jin2002sequential, liu2018survey}), (ii) quality-weighted diversity aiming at task-aware use, and (iii) shape-property joint diversity for tailorable use.

\section{Property Bias: An Example of Lattice Mechanical Metamaterials}
\label{Section1}

\begin{figure}[t]
\centering
\includegraphics[width=0.8\linewidth]{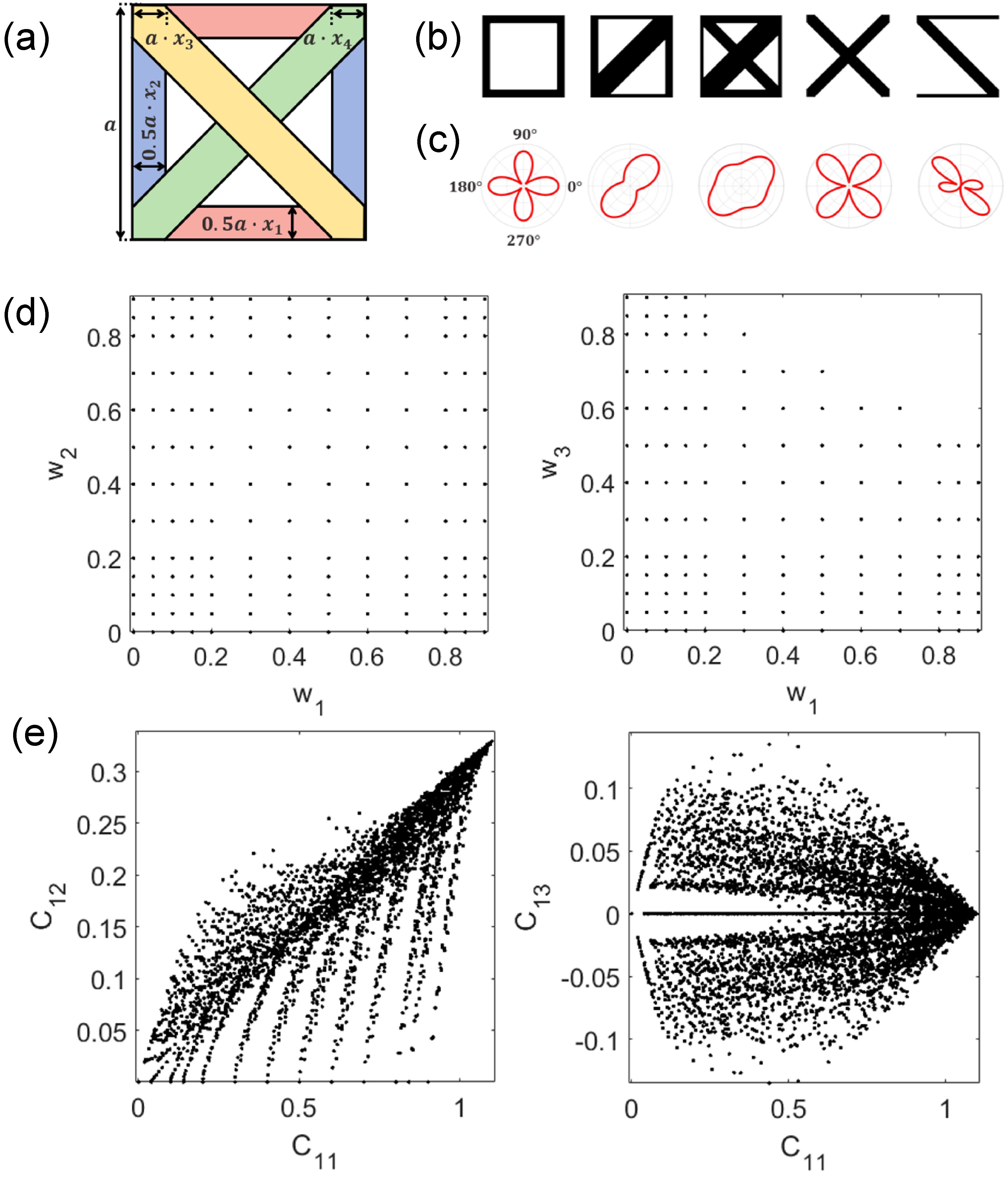}
\caption{Illustration of $\mathcal{D}_{lat}$~\cite{wang2022data}. (a) Microstructure shape representation specific by thickness of each bargroup. (b) Five instances of generated microstructures. (c) The resulting surfaces of homogenized elastic modulus. (d) Data distribution in parametric shape spaces $w_{1}$-$w_{2}$ and $w_{1}$-$w_{3}$. (e) Data distribution in projected property spaces $C_{11}$-$C_{12}$ and $C_{11}$-$C_{13}$.}
\label{fig:d_lat}
\end{figure}

Property bias prevails in existing metamaterial datasets. To convey this point, we examine an example of a lattice-based 2-D mechanical metamaterial dataset. Lattice-based metamaterials have been intensely studied due to their outstanding performance-to-mass ratio, great heat dissipation, and negative Possion's ratio~\cite{yu2018mechanical}. Wang et al. devised a lattice-based dataset~\cite{wang2022data}, to be called $\mathcal{D}_{lat}$ in this work. In the dataset, a unit cell (i.e.,  microstructure or building block) takes six bars aligned in different directions as its geometric primitives (see Figure~\ref{fig:d_lat}(a)). All unit cells can be fully specified by four parameters associated with the thickness of each bar group. The shape generation scheme produces diverse geometric classes (i.e.,  baseline, family, motif, basis, and template), as displayed in Figure~\ref{fig:d_lat}(b). Each class exhibits different topological features, which offer diverse modulus surfaces of homogenized elastic constants $( C_{11}, C_{12}, C_{13}, C_{22}, C_{23}, C_{33} )$ (Figure~\ref{fig:d_lat}(c)).
Figure~\ref{fig:d_lat}(d) shows the nearly uniform sampling in the parametric shape space $ \Omega_w=[0, 1]^4$ used for data population. We removed repeated instances where the entire domain is either solid ($v_f=1$) or void ($v_f=0$); this explains why some regions in Figure~\ref{fig:d_lat}(d) have no data points.

Now we look into the data distribution of $\mathcal{D}_{lat}$. The near-uniform sampling ensures good uniformity in the parametric shape space (Figure~\ref{fig:d_lat}(d)). On the other hand, the corresponding property distributions in Figure~\ref{fig:d_lat}(e) show considerable imbalance, which epitomizes that \textit{data balance in parametric shape space does not ensure the same in property space}. We argue that such property imbalance is prevalent in many metamaterial datasets generated by space-filling design in parametric shape space~\cite{wang2021ih, chan2021metaset, wang2021data, jiang2019global, an2019deep}.
We claim that: (i) metamaterial datasets collected based on naive sampling in parametric shape space are subject to substantial property bias~\cite{wang2021ih, chan2021metaset, wang2021data, jiang2019global, an2019deep}, and more importantly, (ii) this is highly likely to hold true for datasets with generic design representations -- beyond parametric ones -- as well~\cite{chan2021metaset, chan2022remixing, wang2020deep, an2021multifunctional, whiting2020meta}. The general statement is, in part, grounded on the near-zero correlation between shape similarity and property similarity in large-scale metamaterial datasets ($\sim O(10^4)$), consisting of microstructures represented as pixel/voxel, observed by Chan et al~\cite{chan2021metaset}. Overlooking the significant property imbalance, many methods assume that the subsequent stages of DDMD can accurately learn and perform inference under such strong property imbalance, ignoring the compounding impact of data bias~\cite{branco2016survey}.

In addition, diversity alone does not ensure successful deployment of DDMD for design purposes. Imagine a case where a 50k-size dataset with perfect uniformity has been prepared, yet the region associated with a given design task (e.g., high performance-to-mass ratio; high stiffness anisotropy; manufacturability) happens to include a tiny portion of the dataset. This implies, provided a design task has been prescribed, that (i) designers would want to involve the utility of data points for the given task, on top of diversity, during both data acquisition and evaluation; (ii) it could be rather \textit{desirable to promote artificial data bias} towards a certain direction/area associated with the task.

Property bias is inevitable without supervision. Properties -- a function of a given shape -- are unknown before evaluation. Obtaining their values is the major computational bottleneck~\cite{sambasivan2021everyone}, not only at the data preparation stage but also in the whole DDMD pipeline. An undesirable yet prevalent case is: one evaluates all the shape samples with time-consuming numerical analysis (e.g., finite element analysis (FEM); wave analysis) and trains a model on the data, only to end up with a property distribution that is severely biased outside where one had planned to deploy the data-driven model. To circumvent such unwanted scenarios, it is warranted to monitor property distributions at early stages and maneuver the sampling process in a supervised manner \textit{during} data acquisition, not after. As a solution, we propose t-METASET, a task-aware data acquisition framework that tailors data distributions upon user-defined design tasks.


\section{Proposed Method}
\label{Section2}
In this section, we walk readers through the three components of the proposed t-METASET: shape descriptor (Section~\ref{Shape Descriptor}), sparse regressor (Section~\ref{Sparse Regressor}), and diversity-driven sampling (Section~\ref{Diversity-Based Sampling}). Then the algorithm in its entirety is presented. (Section~\ref{The t-METASET Algorithm})
 
\subsection{Shape Descriptor}
\label{Shape Descriptor}


To exploit topologically free variations of building block geometries, metamaterials design often involves a high-dimensional geometric space (e.g., 50$\times$50 pixelated 2-D designs equates a $50^2$-D space). Exploring the vast design space is inefficient and not computationally affordable. Instead we wish to reparameterize instances in the ambient space using a compact yet expressive shape descriptor. The shape descriptor captures essential topological features of metamaterial building blocks and offers a low-dimensional design representation with an acceptable compromise of expressiveness.

In the literature of DDMD, shape descriptors of building blocks roughly fall into three categories: physical descriptors, spectral descriptors, and data-driven descriptors. First, physical descriptors represent a geometry based on geometric features of interest, such as curvature, moment, angle, shape context~\cite{kazmi2013survey}. Hence, the key advantage is high interpretability provided by the physical criteria. For example in DDMD, Chan et al.~\cite{chan2021metaset} employed the division point-based descriptor~\cite{vamvakas2010handwritten}, which recursively identifies centroids of binary images at several granularity levels, and concatenates the coordinates as the descriptor. Second, spectral descriptors exploit finite-dimensional spectral decomposition of ambient shape space. Liu et al.~\cite{liu2020topological} proposed a Fourier transform based descriptor as a topological encoding method for optical metasurfaces. The spectral descriptor enjoys representational parsimony, reconstruction capability (inverse Fourier transform), efficient symmetry handling, and a continuous latent space.
Third, data-driven descriptors exploit data-driven feature engineering. Wang et al.~\cite{wang2020deep} employed a variational autoencoder (VAE)~\cite{kingma2013auto} as a deep generative model for DDMD. It was demonstrated that the latent representation offers a compact shape similarity measure in light of given data, facilitates blending across microstructures, and encodes interpretable geometric patterns.

As a data-driven model involving unsupervised representation learning, VAE learns a compact latent representation that can be used as a shape descriptor~\cite{kingma2013auto}. We advocate the VAE descriptor as the shape descriptor of metamaterial unit cells based on two aspects. First, VAE enjoys the parsimony of a low-dimensional manifold, which is crucial to make a sparse regressor (Section~\ref{Sparse Regressor}) have compact yet expressive predictors, and to expedite the subsequent diversity-driven sampling (Section~\ref{Diversity-Based Sampling}). 
Second, this work also takes advantage of the distributional regularization imposed on the encoder: the latent vectors are enforced to be roughly multivariate Gaussian. The regularization enforces built-in scaling across individual components of the latent representation, rendering diversity-based sampling robust to arbitrary scaling. 

\begin{figure*}[t]
\centering
\includegraphics[width=0.8\linewidth]{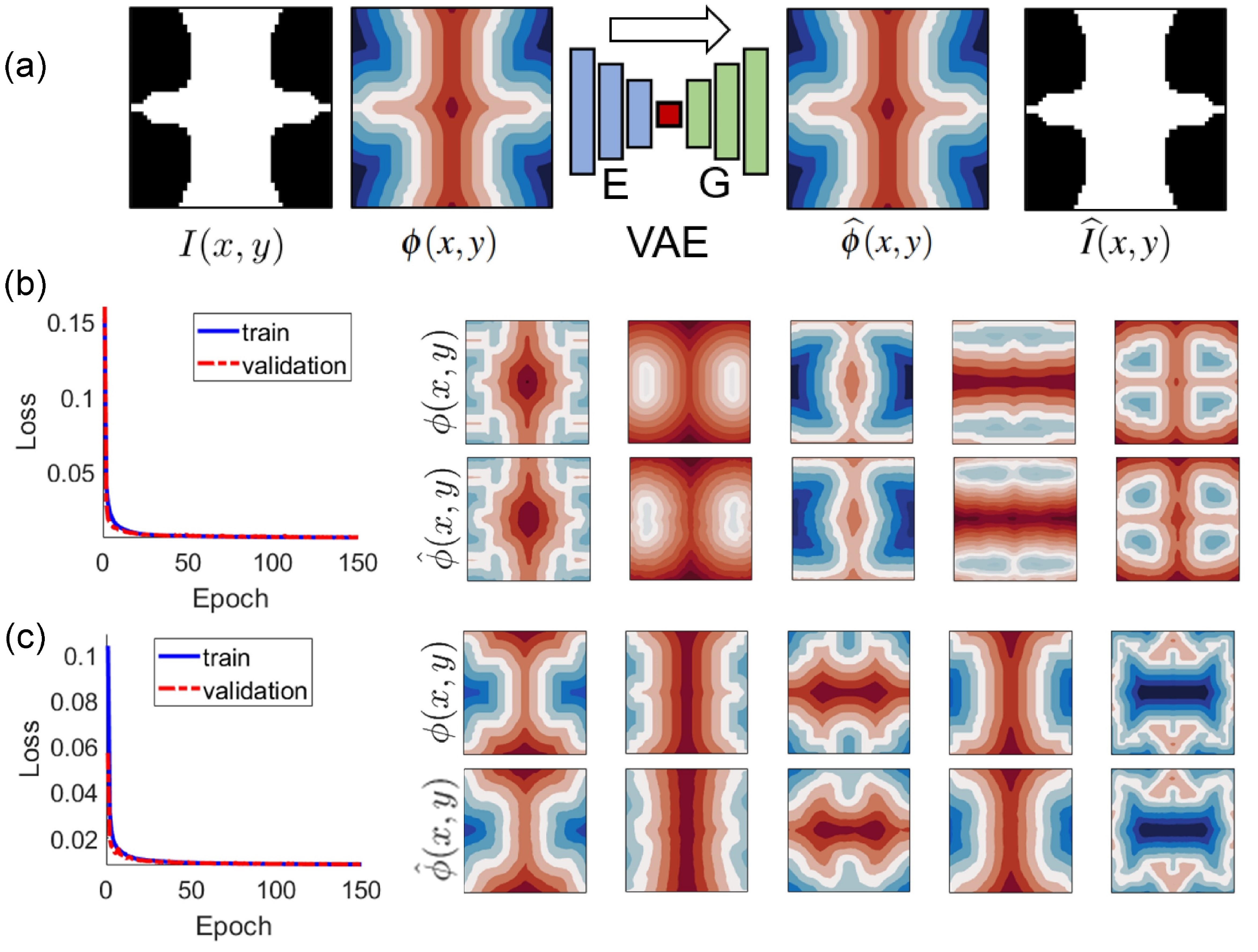}
\caption{Shape VAE. (a) Schematic of the architecture. (b) The training result of $\mathcal{D}_{mix}$. (c) The training result of $\mathcal{D}_{TO}$.}
\label{fig:shapeVAE}
\end{figure*}

Figure~\ref{fig:shapeVAE}(a) depicts the shape VAE used in our study. The VAE involves two key components, encoder $E$ and decoder $G$. Assuming an input instance is given as a discretized image, the encoder involves a set of progressively contracting layers to capture underlying low-dimensional features, until it reaches the bottleneck layer, which provides the latent vector as $\boldsymbol{z}=E(\phi(x, y))$ where $\phi(x, y)$ is the signed distance field (SDF) of a binary microstructure image $I(x, y)$. The decoder, reversely, takes a latent variable from the information bottleneck and generates a reconstructed image as $\widehat{\phi}(x, y)=G(\boldsymbol{z})$. In formatting the shape instances, we prefer the SDF representation to the binary one since (i) SDFs offer richer local information (distance and sign) that unsupervised representation learning can exploit~\cite{dai2017shape}, and (ii) the continuous surface-based representation tends to help generative models produce smoother synthesized instances~\cite{zhang20193d}.

Now we briefly introduce key formulations of VAE. A VAE assumes that given data have come from an underlying random process specified by a latent variable $\boldsymbol{z}$. Each instance $\phi$ and latent variable $\boldsymbol{z}$ are viewed a realization of the conditional distribution $p_{\boldsymbol{\theta}} (\phi|\boldsymbol{z})$ and prior distribution $p_{\boldsymbol{\theta}}(\boldsymbol{z})$, respectively, where $\boldsymbol{\theta}$ is the parameters that specify the distributions. The marginal likelihood of a given instance $\phi$ reads:
\begin{equation}
    \mathrm{log}p_{\boldsymbol{\theta}}(\phi)=KL[q_{\boldsymbol{\psi}}{(\boldsymbol{z}|\phi)}||p_{\boldsymbol{\theta}}{(\boldsymbol{z}|\phi)}]+\mathcal{L}(\boldsymbol{\theta}, \boldsymbol{\psi}; \phi),
\end{equation}
where $KL[\cdot||\cdot]$ is the Kullback-Leibler divergence, a non-negative distance measure between two distributions; $q_{\boldsymbol{\psi}}(\boldsymbol{z}|\phi)$ is the variational posterior that is specified by the paramater $\boldsymbol{\psi}$ and approximates the true posterior $p_{\boldsymbol{\theta}}(\boldsymbol{z}|\phi)$ to bypass the intractability of the marginal distribution~\cite{kingma2013auto}; and $\mathcal{L}(\cdot)$ is the variational lower bound on the marginal likelihood. Usual practice for training is to rearrange the equation and to maximize the evidence lower bound:
\begin{equation}
\begin{split}
    \mathcal{L}(\boldsymbol{\theta}, \boldsymbol{\psi}; \phi)=-KL[q_{\boldsymbol{\psi}}{(\boldsymbol{z}|\phi)}||p_{\boldsymbol{\theta}}{(\boldsymbol{z}|\phi)}]\\+\mathbb{E}_{q_{\boldsymbol{\psi}}{(\boldsymbol{z}|\phi)}}[\mathrm{log}(p_{\boldsymbol{\theta}}(\phi|\boldsymbol{z}))].
\end{split}
\end{equation}
The first right-hand side term $KL[\cdot||\cdot]$ involves the regularization loss that enforces the latent variable $\boldsymbol{z}$ to be distributed as multivariate Gaussian, while the second term denotes the reconstruction loss. The approximated variational lower bound allows stochastic gradient decent to be used for end-to-end training of the whole VAE. For efficient training, standard VAE assumes the prior distribution as $p_{\boldsymbol{\theta}}(\boldsymbol{z})\sim ~\mathcal{N}(\boldsymbol{0}, \boldsymbol{I})$ and the variational posterior as $q_{\boldsymbol{\psi}}(\phi|\boldsymbol{z}')\sim \mathcal{N}(\boldsymbol{\mu}, \boldsymbol{\sigma}^2)$, respectively, where the reparameterization trick~\cite{kingma2013auto} involves a stochastic embedding $\boldsymbol{z}'$ as $\boldsymbol{z}'=\boldsymbol{\mu}+\boldsymbol{\sigma}{\odot} \boldsymbol{\epsilon}$ with a Gaussian noise $\boldsymbol{\epsilon}\sim \mathcal{N}(\boldsymbol{0}, \boldsymbol{I})$. The training reduces to the following optimization problem:
\begin{equation}
    \begin{split}
    \min_{\boldsymbol{\theta}, \boldsymbol{\psi}} [-\mathcal{L}(\boldsymbol{\theta}, \boldsymbol{\psi}; \phi)]=\frac{1}{|\mathcal{S}|}{ \sum p_{\boldsymbol{\theta}}(\phi|\boldsymbol{z}')}\\ 
    -\frac{1}{2} {\sum [1+\mathrm{log}(\boldsymbol{\sigma}^2)-\boldsymbol{\sigma}^2-\boldsymbol{\mu}^2]},
    \end{split}    
    \label{eq:VAE_training}
\end{equation}
where $|\mathcal{S}|$ is the number of shape data.

Figure~\ref{fig:shapeVAE}(b) and (c) report the VAE training results of each dataset, 2-D multiclass blending dataset ($\mathcal{D}_{mix}$)~\cite{chan2022remixing} and 2-D topology optimization dataset ($\mathcal{D}_{TO}$)~\cite{bendsoe2003topology, wang2020data}, respectively. A concise description of the datasets can be found in Section~\ref{data_description}. The VAE architecture was set based on that of Wang et al.~\cite{wang2020deep}. The dimension of the latent space is set as 10, with the trade-off between dimensionality and reconstruction error taken into account. The Adam optimizer~\cite{kingma2014adam} was used to train the VAE with the following setting: learning rate $10^{-4}$, batch size 128, epochs 150, and dropout probability 0.4. Each shape dataset is split into training set and validation set with the ratio of 80\% and 20\%, respectively. In Figure~\ref{fig:shapeVAE}(b) and (c), each training history shows stable convergence behavior for both training and validation. From the plots of SDF instances on the right side, we qualitatively confirm good agreement between the input instances (top) and their reconstruction (bottom), for both training results.

\subsection{Sparse Regressor}
\label{Sparse Regressor}
In t-METASET, a sparse regressor enables active learning and task-aware distributional control under epistemic uncertainty (i.e.,  lack of data). In Section~\ref{GP} we elaborate on why a Gaussian process (GP) is a good choice as the sparse regressor and introduce key formulations of multi-output GPs. Section~\ref{roughness_parameter} details roughness parameters of a GP and how they are harnessed for sampling mode transition in t-METASET.

\subsubsection{Gaussian Processes}
\label{GP}
We implement a GP regressor as the ``agent” of data acquisition in this work. The mission is to learn the underlying structure-property mapping from sparse data, and to pass predictions over unseen shapes as $\widehat{\boldsymbol{p}}=\mathcal{GP}(\boldsymbol{z})$ to batch sequential sampling. In this study, the GP takes the VAE latent shape descriptor as its input, which offers substantial dimension reduction ($50^2$-D $\rightarrow 10$-D in this work). We advocate a GP as the sparse agent due to three key advantages: (i) model parsimony congruent with sparse observations at early stages; (ii) decent modeling capacity of nonlinear structure-property regression (i.e., $\boldsymbol{z}\rightarrow \boldsymbol{p}$);
(iii) roughness parameters as an indicator of model convergence, to be used for sampling mode transition (detailed in Section~\ref{roughness_parameter}).

Building on the advantages of the GP, our novel idea on task-aware property bias control is to (i) construct an \textit{estimated} property similarity kernel $\widehat{L_{\boldsymbol{p}}}$ (Section~\ref{similarity and DPP}) from the GP prediction $\widehat{\boldsymbol{p}}=\mathcal{GP}({\boldsymbol{z}})$, as the counterpart of the shape kernel $\widehat{L_{\boldsymbol{z}}}$, and (ii) employ conditional Determinantal Point Processes (DPP)~\cite{kulesza2012determinantal} -- a probabilistic approach to diversity modeling -- on the estimated property kernel $\widehat{L_{\boldsymbol{p}}}$ to recursively sample a batch based on the expected property diversity. The property kernel $\widehat{L_{\boldsymbol{p}}}$ estimates property similarity, \textit{prior to design evaluation}, not only between train-train pairs, but also train-unseen and unseen-unseen ones. In this way, the sampler of t-METASET recommends a batch $\mathcal{B}$ hinging on both estimated property diversity and shape diversity. It is important to note that, at an incipient phase, we do not rely on $\widehat{L_{\boldsymbol{p}}}$, as the predictive performance of a multivariate multiresponse GP ($\mathbb{R}^{D_{z}} \rightarrow \mathbb{R}^{D_{\boldsymbol{p}}} $, where $D_{\boldsymbol{p}}$ is the property dimensionality) trained on tiny data is not reliable. We determine the turning point -- when to start to respect the GP prediction -- based on the convergence history of a set of the GP hyperparameters: roughness parameters (i.e., scale parameters). 

As a background to the roughness parameters, we introduce key formulations of GPs. A GP is a collection of random variables, any of whose finite subset is distributed as multivariate Gaussian~\cite{rasmussen2003gaussian}. Given a set of observations, a GP with $D_{\boldsymbol{p}}$ responses is fully specified by its mean and covariance functions as 
\begin{equation}
    f\sim\mathcal{GP}({\boldsymbol{\mu}}(\boldsymbol{z}), cov(\boldsymbol{z},\boldsymbol{z}')),
\end{equation}
where $\boldsymbol{\mu}(\cdot)$ is the mean function; $cov(\cdot,\cdot)$ is the covariance function; $f$ is a function viewed as a realization from the underlying distribution. For the multivariate input $\boldsymbol{z}$ and the multiresponse outputs $\boldsymbol{p}$ in our study, the covariance function reads: $cov(\boldsymbol{z},\boldsymbol{z}')={\boldsymbol{\Sigma}}\otimes r(\boldsymbol{z},\boldsymbol{z}'),$
where $\boldsymbol{\Sigma}$ is the $D_{\boldsymbol{p}}\times D_{\boldsymbol{p}}$ dimensional multiresponse prior variance, $\otimes$ is the Kronecker product, and $r(\cdot, \cdot)$ is the correlation function. In this work we use the squared exponential correlation function given as
\begin{equation} \label{eq:correlation}
r(\boldsymbol{z}, \boldsymbol{z}')=\text{exp}(-(\boldsymbol{z}-\boldsymbol{z}')\Theta(\boldsymbol{z}-\boldsymbol{z}'))
\end{equation}
where $\Theta=\text{diag}(10^{\boldsymbol{\omega}})$ and $\boldsymbol{\omega}=[\omega_1, \cdots, \omega_{D_z}]^T$ is the vector of roughness parameters~\cite{bostanabad2019globally}. Given a dataset $\mathcal{D}={\{ (({\boldsymbol{z}}_1, \ldots, {\boldsymbol{z}}_{D_z}), ({\boldsymbol{p}}_1, \ldots, {\boldsymbol{p}}_{D_{\boldsymbol{p}}})) \}}_{i=1}^{n}$, a point estimate of the hyperparameters can be found through maximizing the Gaussian likelihood function:
\begin{equation}
\begin{split}
[\widehat{\boldsymbol{\beta}}, \widehat{\boldsymbol{\Sigma}}, \widehat{\boldsymbol{\omega}}]=\underset{[\boldsymbol{\beta}, \boldsymbol{\Sigma},\boldsymbol{\omega}]}{\arg\min} [\frac{n}{2}\text{log}(\det({\boldsymbol{\Sigma}}))+\frac{1}{2}\text{log}(\text{det(}\boldsymbol{R}))
+\frac{1}{2\sigma^2}(\boldsymbol{p}-\boldsymbol{1}\boldsymbol{\beta})^T{\boldsymbol{R}}^{-1}(\boldsymbol{p}-\boldsymbol{1}\boldsymbol{\beta})]
\end{split}
\end{equation}
where $\boldsymbol{1}$ is an $n\times D_{\boldsymbol{p}}$ dimensional vector of ones, ${\boldsymbol{\beta}}=[\beta_1,\ldots,\beta_{D_{\boldsymbol{p}}}]^T$ is $1\times {D_{\boldsymbol{p}}}$ dimensional vector of weights,  $\text{log}(\cdot)$ is the natural logarithm, $\boldsymbol{R}$ is the $n \times n$ correlation matrix with $(i, j)$-${th}$ element $R_{ij}$ given as $r({\boldsymbol{z}}_i, {\boldsymbol{z}}_j)$ for $i, j =1,\ldots, n$, and  det($\cdot$) is the matrix determinant operation. In the above formulation of the likelihood we have assumed a constant prior mean function as ${\boldsymbol{1}}{\boldsymbol{\beta}}$. More complex basis functions can be used to represent the prior mean (e.g., linear, or quadratic); however, this is not advised as this information is typically not known \textit{a priori}, and is likely to compromise model accuracy when chosen incorrectly. 

After approximation of the hyperparameters, the posterior predictive distribution for an unobserved input $\boldsymbol{z}_{new}$ can be obtained by conditioning the prior distribution on the observed data $\mathcal{D}$ \cite{van2020integration}. Specifically, the mean and the covariance of the posterior predictive distribution is given as
\begin{align}
\begin{split}
    {\boldsymbol{\mu}}(\boldsymbol{z}_{new}) & = \hat{\boldsymbol{\beta}}+\boldsymbol{r}^T(\boldsymbol{z}_{new})\boldsymbol{R}^{-1}(\boldsymbol{p}-\boldsymbol{1}\hat{\boldsymbol{\beta}}),\\
    cov(\boldsymbol{z}_{new}) & = \hat{\boldsymbol{\Sigma}}[r({\boldsymbol{z}}_{new}, \boldsymbol{z}_{new}) - \boldsymbol{r}^T({\boldsymbol{z}}_{new})\boldsymbol{R}^{-1}\boldsymbol{r}({\boldsymbol{z}}_{new})+\boldsymbol{W}^T(\boldsymbol{1}^T\boldsymbol{R}^{-1}\boldsymbol{1})\boldsymbol{W} ],
\end{split}
\end{align}
where $\boldsymbol{r}(\boldsymbol{z}_{new})$ is an $n \times 1$ dimensional vector whose $i-{th}$ element is given as $r(\boldsymbol{z}_{new}, z_i)$ for $i=1,\ldots,n$, $\boldsymbol{W}=\boldsymbol{1}^{\prime} - \boldsymbol{1}^T\boldsymbol{R}^{-1}\boldsymbol{r}(\boldsymbol{z}_{new})$, and $\boldsymbol{1}^{\prime}$ is a $D_{\boldsymbol{p}} \times 1$ dimensional vector of ones.

\subsubsection{Roughness parameters}
\label{roughness_parameter}
Informally speaking, roughness parameters $\boldsymbol{\omega}=[\omega_1, \ldots \omega_{D_z}]^T$ dictate fluctuation levels of responses w.r.t. each predictor (each component of $\boldsymbol{z}$ in our study), in light of given data. Bostanabad et el.~\cite{bostanabad2019globally} used the fluctuations of roughness parameters with Eq.~\ref{eq:correlation} and their estimated variance to qualitatively determine if sufficient samples were collected during GP training. Building on that, we monitor the roughness parameters $\boldsymbol{\omega}$ and take the convergence of roughness parameters as a proxy for model convergence. The roughness residual serves as the transition criterion across sampling modes. We define the convergence criterion involving the roughness residual metric $\Delta$ as follows:
\begin{equation}
\begin{split}
\Delta^{(t+1)}=\sqrt{\frac{1}{D_z}\||\boldsymbol{\omega}^{(t+1)}-{\boldsymbol{\omega}}^{(t)}\||^2}
\leq \tau \quad (t=1, \ldots, t_{max})
\label{eq:roughness}~
\end{split}
\end{equation}
where $\tau$ is a threshold associated with the sampling mode transition. At an early stage the roughness residual exhibits a ``transient” behavior. As a stream of data comes in, the residual converges to zero, implying a mild convergence of the GP. In this work, we set two different values of threshold namely, $\tau_1$ and $\tau_2$ where $\tau_1 > \tau_2$. We assume each convergence criterion is met if the residuals of five consecutive iterations are below the threshold. $\tau_1$ is to identify a mild convergence, indicated by the larger tolerance. Once met, t-METASET initiates Stage II, where estimated property diversity serves as the main sampling criterion. Meanwhile, the smaller threshold $\tau_2$ is used to decide when to stop the GP update: as the size of training data accumulates, the variations of roughness parameters get unnoticeable~\cite{bostanabad2019globally}, whereas the computational cost of fitting the GP rapidly increases as $\sim O(|\mathcal{D}^{(t)}|^3)$ due to the inversion of covariance matrix $\boldsymbol{R}$. We prioritize speed, at the modest cost of prediction accuracy. Detailed implementation with the other pillars can be found in Section~\ref{The t-METASET Algorithm}. When reporting the results of t-METASET, we will include the history of the residuals, in addition to that of diversity metrics.

\subsection{Diversity-Based Sampling}
In this section, we elaborate on diversity-based batch sequential sampling. It maneuvers the data acquisition, leveraging both the compact shape descriptor distilled by the VAE (Section~\ref{Shape Descriptor}) and iteratively refined prediction offered by the GP agent (Section~\ref{Sparse Regressor}), from beginning to end of t-METASET. Recalling the mission of t-METASET -- task-aware generation of balanced datasets -- we advocate DPP-based diversity sampling primarily based on three key advantages:
(i) DPPs offer a variety of practical extensions (e.g., cardinality constraint, conditioning) that facilitate the active learning of t-METASET;
(ii) The probabilistic modeling from DPP captures the trade-off between diversity and quality;
(iii) Importantly, DPPs are flexible in terms of handling distributional characteristics in that most object-driven sampling approaches~\cite{snoek2012practical} support either exploration (diversity of input) or exploitation (quality of output), while DPPs do all the combinations of diversity (input/output) and quality (shape/property/joint) without restrictions.

t-METASET builds on a few extensions of DPPs.
Section~\ref{similarity and DPP} provides fundamental concepts related to DPP. Section~\ref{conditional DPP} introduces conditional DPPs that are key for DPP-based active learning, and brings up the scalability issue of massive similarity kernels. As a workaround, a large-scale kernel approximation scheme is introduced in  Section~\ref{low-rank kernel approximation}. Section~\ref{Quality-Weighted Diversity} addresses how to accommodate design quality into DPP, which enables ``task-aware” dataset construction.

\label{Diversity-Based Sampling}
\subsubsection{Similarity and DPP}
\label{similarity and DPP}
In general, an instance of interest could be represented as a vector $\mathbf{x}$. A similarity metric between items $i$ and $j$ can then be quantified as a monotonically decreasing function of the distance in the virtual item space as
\begin{equation}
s_{ij}=T(h(\boldsymbol{x}_i, \boldsymbol{x}_j)),
\label{eq: similarity}
\end{equation}
where $s_{ij}$ is the pairwise similarity between items $i$ and $j$, $h(\cdot, \cdot)$ is a distance function, $T$ is a monotonically decreasing transformation (i.e., the larger a distance, the smaller the similarity is).
One way to represent all the pairwise similarities of a given set is to construct the $n{\times}n$ similarity matrix $L$ as $L_{ij}=s_{ij}$, where $n=|L|$ is the set cardinality (i.e., dataset size). The matrix is often called a \textit{similarity kernel} in that it converts a pair of items into a distance measure (or a similarity measure, equivalently). While any combinations of similarity and transformation are supported by the formalism above, usual practice favors transformations that result in positive semi-definite (PSD) kernels for operational convenience, such as matrix decomposition. Following this, we employ Euclidean distance $h(\boldsymbol{x}_i, \boldsymbol{x}_j)=\sqrt{|| \boldsymbol{x}_i-\boldsymbol{x}_j ||^2}$ and the square exponential transformation. The resulting similarity kernel reads:
\begin{equation}
L_{ij}= \text{exp}\left(\frac{-{||{\boldsymbol{x}_i-\boldsymbol{x}_j}||^2}}{2\sigma_{L}^2}\right),
\label{eq: Gaussian similarity}
\end{equation}
where $\sigma_L$ is a length-scale parameter (i.e.,  bandwidth) that tunes the correlation between items.

DPPs provide an elegant probabilistic modeling that favors a subset comprised of diverse instances~\cite{kulesza2012determinantal}. They have been employed for a variety of applications that take advantage of set diversity, such as recommender systems~\cite{gartrell2016bayesian}, summarization~\cite{chao2015large}, object retrieval~\cite{affandi2014learning}. The defining property of DPPs is:
\begin{equation}
p(X=\mathcal{A}){\propto}{\mathrm{det}}(L_{\mathcal{A}}),
\end{equation}
where $L_{\mathcal{A}}$ is a subset of a ground set $L=L_{\mathcal{V}}$ indexed by $\mathcal{A}$, and $p(X=\mathcal{A})$ is the probability to sample $\mathcal{A}$. The property has an intuitive geometric interpretation: det($\mathcal{A}$) is associated with the hypervolume spanned by the constituent instances. If the catalog $\mathcal{A}$ includes any pair of items that is almost linearly dependent on each other, the corresponding volume would be nearly zero, making $\mathcal{A}$ unlikely to be selected. De-emphasizing such cases, the DPP-based sampling serves as a subset recommender that favors a subset of diverse items. In this study, we set the batch size $k$ to be constant at $k=10$ using $k$-DPP~\cite{kulesza2011k} as follows:
\begin{equation}
p(X=\mathcal{B})=\frac{\mathrm{det}(L_\mathcal{B})}{ \sum_{|\mathcal{B}'|=k} \mathrm{det}(L_{\mathcal{B}'})},
\label{eq:k-dpp}
\end{equation}
where $L_\mathcal{B}$ denotes a submatrix indexed by the items that constitute a batch, or subset, $\mathcal{B}\in \mathcal{V}$.

\subsubsection{Conditional DPP}
\label{conditional DPP}
Our data acquisition grows a dataset using active learning. At each iteration, the similarity kernels should be recursively updated so that sampling a new batch leverages the latest information of all evaluated observations. This enables the sampler to (i) avoid drawing duplicate samples that have been observed, and to (ii) promote samples that are diverse not only within a given batch, but \textit{across} a sequence of batches~\cite{affandi2012markov}. In DPP, such a kernel update is supported via conditioning a DPP on the instances observed so far. DPPs are closed under conditioning operations; i.e.,  a conditional DPP is also a DPP~\cite{borodin2005eynard, affandi2012markov, Gartrell2017low}. This implies that DPP-based sampling can be iteratively applied to similarity kernels to achieve across-batch diversity, as well as within-batch diversity~\cite{affandi2012markov}.
Let $\mathcal{B}$ and $\mathcal{V}$ be the batch and the ground set at the $i$-th iteration, respectively. Given the DPP kernel $L^{(i)}$ at that iteration, a recursive formula for the conditional kernel $L^{(i+1)}$ reads:  
\begin{equation}
L^{(i+1)}=\left({\left(L^{(i)}+I_{\overline{\mathcal{B}}}\right)}^{-1}_{\overline{\mathcal{B}}}\right)^{-1}-I,
\label{eq:conditioning}
\end{equation}
where $\overline{\mathcal{B}}=\mathcal{V}\setminus \mathcal{B}$. Due to the cascaded matrix inversions involving cubic time complexity, the equation does not scale well to the large-scale kernels with instances $\sim O(10^4)$. Furthermore, t-METASET demands at least a few hundreds of conditioning. Even just storing a $88,180^2$-size similarity kernel for $\mathcal{D}_{TO}$ with double precision takes up about 62 gigabytes. In brief, Eq.~\ref{eq:conditioning} is intractable for large-scale similarity kernels of our interest. 

\subsubsection{Large-Scale Kernel Approximation}
\label{low-rank kernel approximation}
To circumvent the scalability issue, we leverage large-scale kernel approximation~\cite{rahimi2007random}. Recalling that we have employed the Gaussian similarity kernel (Section~\ref{similarity and DPP}), we harness the shift-invariance (i.e.,  $L(\boldsymbol{x}, \boldsymbol{y})=L(\boldsymbol{x}-\boldsymbol{y})$) by implementing random Fourier feature (RFF)~\cite{rahimi2007random} as an approximation method. It builds on the Bochner theorem~\cite{rudin2017fourier}, which states that the Fourier transform $\mathcal{F}$ of a properly scaled shift-invariant (i.e.,  stationary) kernel $L$ is a probability measure $p(f)$ as follows:
\begin{equation}
L(\boldsymbol{x}-\boldsymbol{y})= \int_{\Omega}^{} p(f)\text{exp}(jf'(\boldsymbol{x}-\boldsymbol{y})) \,df ,
\end{equation}
where $j$ is the imaginary unit $\sqrt{-1}$, $p(f)=\mathcal{F}[L(\boldsymbol{x}-\boldsymbol{y})]$ is the probability distribution, $D_{V}$ ($\ll n$) is the feature dimension, and $ \boldsymbol{x}, \boldsymbol{y} \in \Omega$. By setting $\zeta_f(\boldsymbol{x})=\text{exp}(jf'\boldsymbol{x})$, we recognize that $L(\boldsymbol{x}, \boldsymbol{y})=\mathbb{E}_f[\zeta_f(\boldsymbol{x})\zeta_f(\boldsymbol{y})^{*}]$, implying that $\zeta_f(\boldsymbol{x})\zeta_f(\boldsymbol{y})^{*}$ is an unbiased estimate of the kernel to be approximated. The estimate variance is lowered by concatenating $D_{V}$ ($\ll n$) realizations of $\zeta_f(\boldsymbol{x})$. For a real-valued Gaussian kernel $L$, the probability distribution $p(f)$ is also Gaussian, and $\zeta_f(\boldsymbol{x})$ reduces to cosine. Under all the considerations so far, the $D_V\times n$ RFF becomes:
\begin{equation}
V(\boldsymbol{x}) = \sqrt{\frac{2}{D_{V}}}[\mathrm{cos}(f_1'\boldsymbol{x}+b_1), \ldots, \mathrm{cos}(f_{D_V}'\boldsymbol{x}+b_{D_V})]^T,
\label{RFF}
\end{equation}
where $\{f_1, \ldots, f_{D_{V}} \} \stackrel{iid}{\sim} \mathcal{N}(0, 1)$ and ${b_1, \ldots, b_{D_V}\}}\stackrel{iid}{\sim} \mathcal{U}[0, 2\pi]$.
Given an RFF $V$, the updated feature $V'$ conditioned on a batch $\mathcal{B}$ has the following closed-form expression~\cite{Gartrell2017low}:
\begin{equation}
V'=V_{\overline{\mathcal{B}}}Z^{\mathcal{B}}( I-V^T_{\mathcal{B}}(V_{\mathcal{B}}V^T_{\mathcal{B}})^{-1}V_{\mathcal{B}} ),
\label{eq:low_rank_markov}
\end{equation}
where the true kernel can be estimated via $L \approx V'(V')^T $. Now the matrix inversions become amenable as the time complexity decreases to $O(|\mathcal{B}|^3)$ with $|\mathcal{B}|=k\ll n$.

\subsubsection{Quality-Weighted Diversity for Task-Aware Sampling}
\label{Quality-Weighted Diversity}
Lastly, we take into account user-defined quality, in addition to diversity, to construct datasets that are not only balanced but also \textit{task-aware}.
This study is dedicated to \textit{pointwise} design quality, where a pointwise $n \times 1$ quality vector $q(\mathbf{z}, \widehat{\mathbf{p}})$ associated with a design task serves as an additional weight to a feature $V'$. The resulting feature $D_V \times n$ matrix $V''$ reads
\begin{equation}
V''=[\overbrace{q(\boldsymbol{z},\widehat{\boldsymbol{p}}^{(t)}) \dots q(\boldsymbol{z},\widehat{\boldsymbol{p}}^{(t)})}^{D_V}]^T \circ  V',
\label{eq:rff_final}
\end{equation}
where $\circ$ denotes the Hadamard product (i.e., elementwise multiplication).

The quality-weighted DPP sampling could seem similar to Bayesian optimization (BO)~\cite{snoek2012practical} in that: (i) quality contributes to exploitation given design attributes of interest, whereas diversity supports exploration, and (ii) both use sequential sampling, taking a GP as the surrogate. We highlight their differences as: (i) t-METASET does not take the uncertainty provided by the GP regressor – at least under the current setup – as a sampling criterion; (ii) diversity is the main driver of the sequential DPP sampling, whereas in BO, exploration (diversity) is ultimately a means for exploitation (quality); (iii) t-METASET is primarily driven by \textit{pairwise} DPP kernels, taking a pointwise quality as an option, whereas BO is driven by a \textit{pointwise} acquisition function;  (iv) t-METASET handles quality that accommodates distributional attributes of shape, property, and even the combination of them, while for BO no acquisition functions have been proposed that explicitly consider property distribution; (v) t-METASET has more flexibility in terms of tailoring distributional characteristics, while standard BO ends up biasing both shape and property distributions to reach the global optimum of a black-box cost function. Quantitative comparisons between t-METASET and BO would be an interesting topic but is currently beyond the scope of this work, as t-METASET can only downsample out of $|\mathcal{S}|$ \textit{finite} points in the VAE latent space, whereas standard BO takes \textit{infinitely many} continuous inputs into account. The validation would be viable under the following extensions: (i) the decoder of the VAE joins the t-METASET algorithm to generate new shapes $\widehat{\phi}(x, y)=G(\boldsymbol{z})$, not existing in the given shape dataset $\mathcal{S}$, and (ii) continuous DPP~\cite{affandi2013approximate} can be employed to recommend diverse samples from a \textit{continuous} landscape, learned from the discrete data points provided by users. This is our future work.
\subsection{The t-METASET Algorithm}
\label{The t-METASET Algorithm}
\begin{figure*}[t]
\centering
\includegraphics[width=1.00\linewidth]{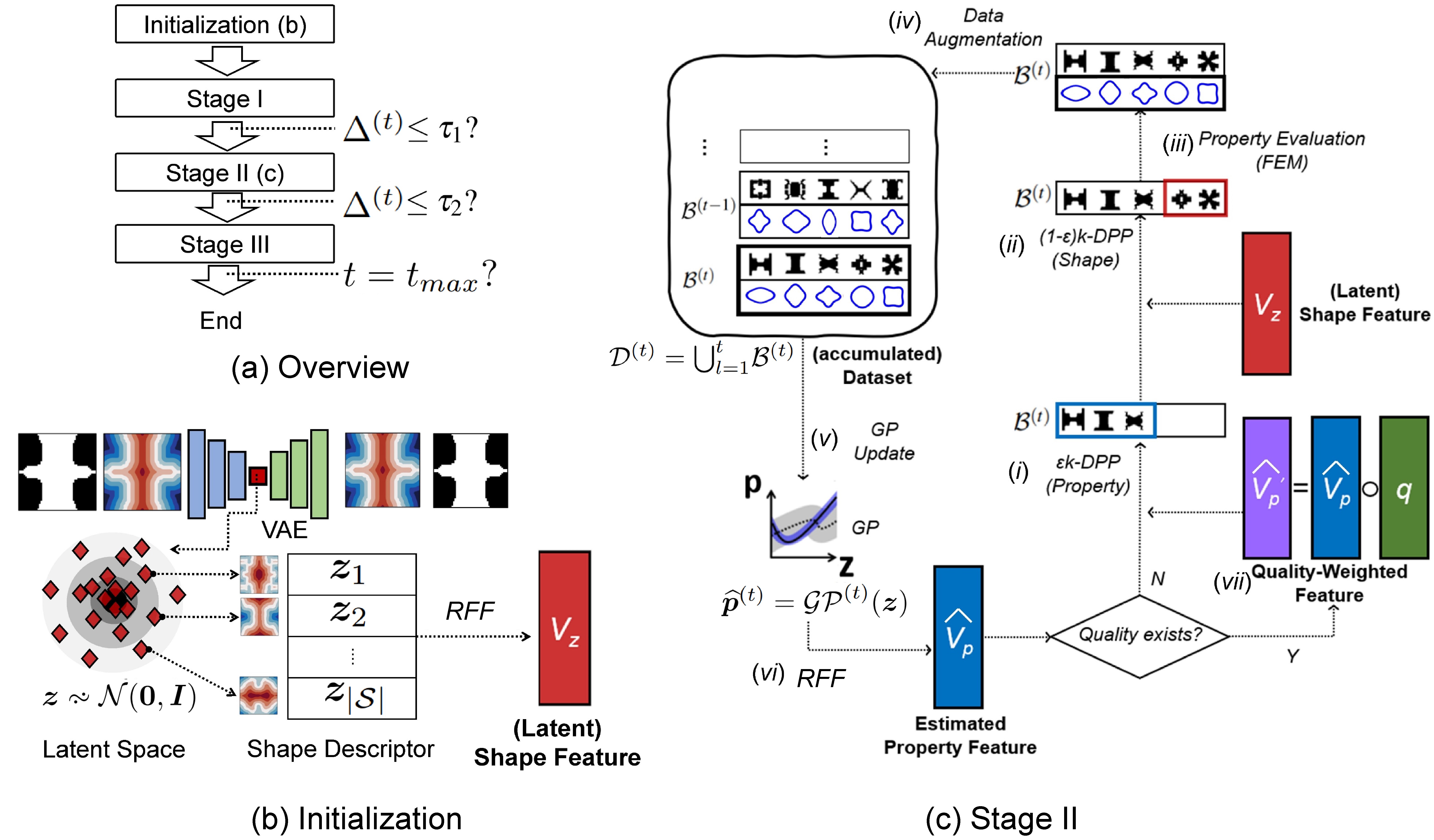}
\caption{A visual overview of t-METASET. (a) Sampling mode transition associated with the roughness residual $r^{(t)}$. (b) Initialization. The VAE is trained on the given shape-only dataset $\mathcal{S}$. The latent variables are roughly distributed as multivariate Gaussian. The latent representation is taken as the shape descriptor, whose concatenation forms the $|\mathcal{S}| \times D_z$ matrix, where $|\mathcal{S}|$ is the shape set cardinality ($\sim O(10^4)$). RFF follows to extract a $|\mathcal{S}| \times D_V$-sized feature of shape feature to be used for the DPP sampling based on shape diversity. (c) A simplified flowchart of Stage II. Details are stated in the main body. Stage I shares the structure as Stage II but is driven only by shape feature (no step (iii)). Stage III is equivalent to Stage II except for the GP update (step (v)). The proposed data acquisition ends when user-defined termination criteria are met (e.g., maximum iteration).}
\label{fig:flowchart}
\end{figure*}

In this section, we detail how to seamlessly integrate the three main components introduced: (i) the latent shape descriptor from the shape VAE, (ii) a sparse regressor as the start-up agent, and (iii) the batch sequential DPP-based sampling that suppresses undesirable bias while enforcing an intentional one. Visual illustration of t-METASET is presented in Figure~\ref{fig:flowchart}. Figure~\ref{fig:flowchart}(a) shows a flow of t-METASET, whose transition is determined by the roughness residual of the GP agent. Given a shape-only, Figure~\ref{fig:flowchart}(b) depicts the initialization of t-METASET supported by VAE shape descriptor~\ref{Shape Descriptor} and large-scale kernel approximations~\ref{low-rank kernel approximation}. The key sampling procedure of t-METASET is illustrated in Figure~\ref{fig:flowchart}(c).

\subsubsection{Initialization}
Figure~\ref{fig:flowchart}(b) illustrates the initialization of t-METASET, which involves VAE training, latent shape descriptor, and RFF extraction from the descriptor. The framework takes the following input arguments: the shape-only dataset $\mathcal{S}$ comprised of SDF instances $\phi(x, y)$, batch cardinality $k$, the ratio of property samples in each batch $\epsilon$, and optionally a pointwise quality function $q(\boldsymbol{z}, \widehat{\boldsymbol{p}})$ that reflects a design task if declared in advance. A shape VAE is trained on $\mathcal{S}$ with the dimension of latent space $D_z$, which is 10-D herein (Figure~\ref{fig:flowchart}(b)).

Then we draw the $D_V\times n$-sized RFF $V_{\boldsymbol{z}}$ (\ref{RFF}) of the $n \times n$ shape similarity kernel, $L_{{\boldsymbol{z}}}$. This feature is to be recursively updated by conditioning on a series of collected batches. For initialization of conditional DPPs over the shape feature, we follow the procedure of Affandi et al.~\cite{affandi2012markov}.

\subsubsection{Stage I}
During Stage I, the GP model's roughness parameter $\boldsymbol{\omega}$ shows large fluctuation due to lack of data. The sampling only relies on shape diversity, because the property prediction of the GP given unseen latent variables is not reliable yet. This stage also can be viewed as initial exploration driven by the pairwise shape dissimilarity -- as an analog to initial passive space-filling design -- where $|\mathcal{D}|\sim O(10^4)$ discrete data points are given as a pool for sampling.

\subsubsection{Stage II}
Figure~\ref{fig:flowchart}(c) provides an overview of Stage II -- the core sampling stage of t-METASET. As more data come in, the roughness residual $\Delta^{(t)}$ (Eq.~\ref{eq:roughness}) approaches zero and becomes stable. Provided that the roughness residual falls under the first threshold $\tau_1$  for five consecutive iterations, the t-METASET framework assumes that the GP prediction is ready to be appreciated. t-METASET proceeds to the next sampling phase Stage II, where t-METASET harnesses the \textit{estimated} property diversity, in addition to shape diversity, as the main criterion. The key is to introduce the RFF of the \textit{estimated} property $\widehat{V_{\boldsymbol{p}}} $, building on the GP prediction $ \widehat{{\boldsymbol{p}}}=\mathcal{GP}({\boldsymbol{z}})$.

Now we detail each step described in Figure~\ref{fig:flowchart}(c).
(i) Given a ratio of property samples $\epsilon$ in a given batch, the DPP sampler draws $ \epsilon k\in \mathbb{N}$ instances from the property RFF $V_{\boldsymbol{p}}$ based on property diversity, weighted by task-related quality when a task is specified. (ii) The rest of the batch is filled by $ (1-\epsilon) k$ samples from the shape RFF, to complement possible lack of exploration in the shape descriptor space $\Omega_{\boldsymbol{z}}$. Herein, the shape RFF must be updated with respect to batch $\mathcal{B}_{\epsilon}$ first, to reflect the latest information. Once sampled, the shape feature is updated again with respect to the rest of the shapes in $\mathcal{B}_{1-\epsilon}$ just selected, for the next iteration. (iii) The microstructures of the batch are observed by design evaluation -- FEM with energy-based homogenization~\cite{xia2015design, andreassen2014determine} in this study -- to obtain the true properties (e.g., $\boldsymbol{p}=\{C_{11}, C_{12}, C_{22}\}$). (iv) The true properties replace the GP prediction in the given batch $\mathcal{B}^{(t)}$. (v) Then the evaluated batch updates the GP to refine the property prediction as $\widehat{{\boldsymbol{p}}}^{(t)}=\mathcal{GP}^{(t)}({\boldsymbol{z}})$ for the next iteration.  (vi) The refined prediction demands the update of a new property RFF, as well as the conditioning of it on the entire dataset $\mathcal{D}^{(t)} = {\bigcup}_{t=1}^{t_{max}} \mathcal{B}^{(t)} $ collected so far. (vii) If a quality function $q({\boldsymbol{z}}, \widehat{{\boldsymbol{p}}})$ over design attributes has been specified, it can be incorporated into the latest property RFF by invoking Eq.~\ref{eq:rff_final} to prompt a task-aware dataset.

\subsubsection{Stage III}
Stage III shares all the settings of Stage II except for the GP update. The main computational overhead of Stage II comes from GP fitting as it involves matrix inversion with the time complexity $\sim O(|\mathcal{D}^{(t)}|^3)$. To bypass the overhead, we stop updating the GP if the roughness residual falls under $\tau_2$ for five consecutive iterations.
During Stage III, our algorithm can quickly identify diverse instances from a large-scale dataset ($\sim O(10^4)$), without the scalability issue. The main product of t-METASET is a high-quality dataset $\mathcal{D}^{t_{max}}=\bigcup_{t=1}^{t_{max}} \mathcal{B}^{(t)}$, which is not only diverse but task-aware. 

\section{Results}
In this section, the results of t-METASET are presented. As benchmarks, the two large-scale mechanical metamaterial libraries~\cite{chan2022remixing, wang2020data} are used for validation. Data description on the two datasets is provided in Section~\ref{data_description}. 
We propose an interpretable diversity metric in Section~\ref{distance_gain} for fair evaluation of t-METASET.
To accommodate various end-uses in DDMD, we validate t-METASET under three hypothetical deployment scenarios: (i) \textit{diversity only} for generic use (balanced datasets; Section~\ref{diversity_only}), (ii) \textit{quality-weighted diversity} for particular use (task-aware datasets; Section~\ref{Scenario II: Quality-Weighted Diversity}), and (iii) \textit{joint diversity} for tailorable use (tunable datasets; Section~\ref{Scenario III: Joint Diversity}). Basic settings include: batch cardinality as $k=10$; property sample ratio during Stage II as $\epsilon=0.8$; the RFF size as $D_V=3,000$; maximum iteration as $t_{max}=500$; first and second threshold of roughness parameters as $\tau_1=0.02$ and $\tau_2=0.01$, respectively; iteration tolerance of roughness convergence as $i_{tol}=5$. Lastly, we focus on producing datasets with sizes of either $3,000$ or $5,000$ (i.e., $t_{max}=300$ or $500$, respectively). 

\subsection{Datasets}
\label{data_description}
\begin{figure}[t]
\centering
\includegraphics[width=0.60\linewidth]{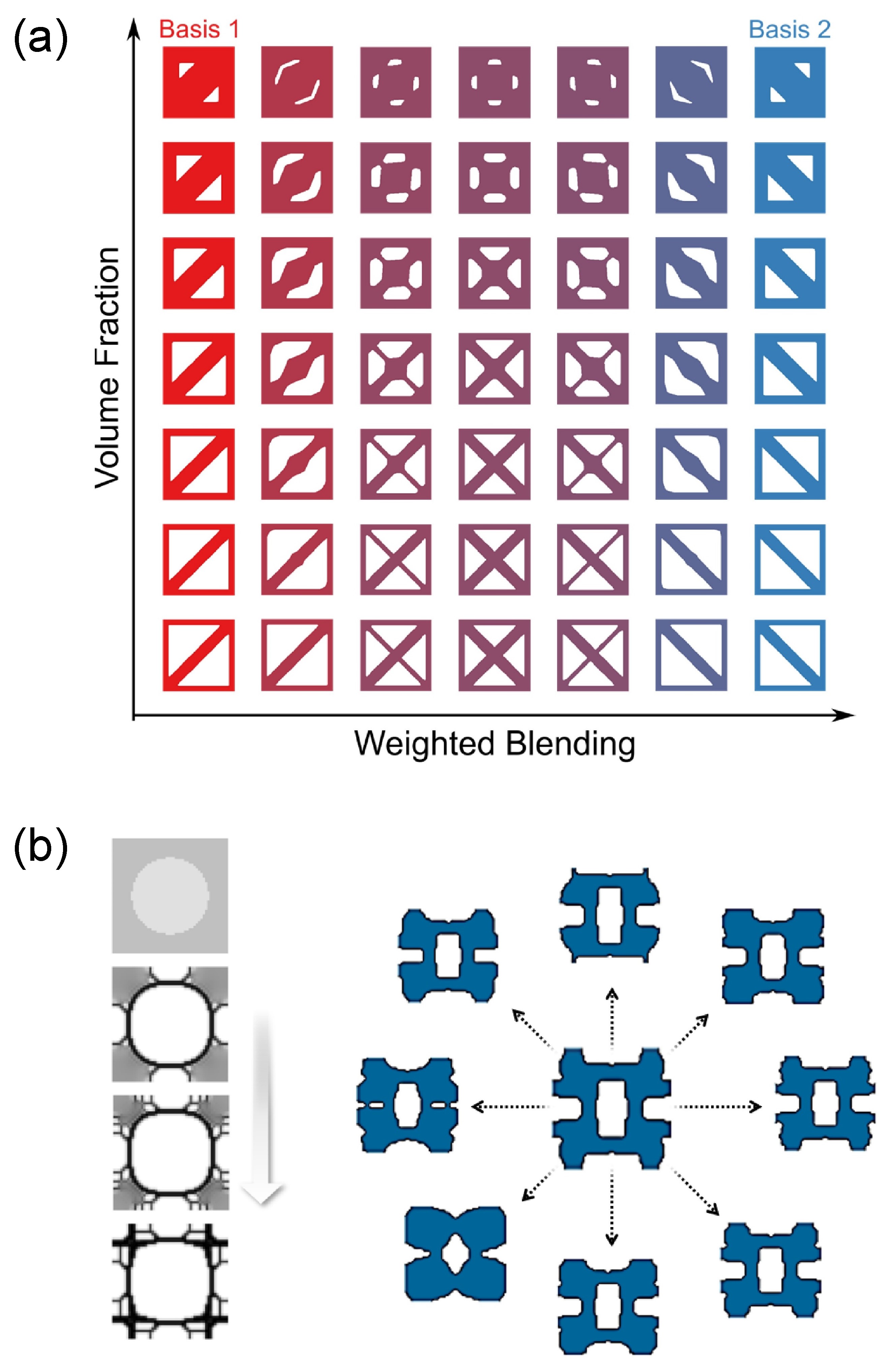}
\caption{Illustration on shape generation schemes of each dataset. (a) $\mathcal{D}_{mix}$~\cite{chan2022remixing}: example of blending the SDFs of basis shapes and varying their volume fractions to produce new unit cells. (b) $\mathcal{D}_{TO}~$~\cite{wang2020data}: (left) an example of design evolution by inverse topology optimization with respect to a target property; (right) the stochastic shape perturbation applied to a given microstructure. } 
\label{fig:data_description}
\end{figure}

We introduce two mechanical metamaterial datasets, in addition to $\mathcal{D}_{lat}$, to be used for validating t-METASET: (i) 2-D multiclass blending dataset ($\mathcal{D}_{mix}$)~\cite{chan2022remixing}, and (ii) 2-D topology optimization dataset ($\mathcal{D}_{TO}$)~\cite{wang2020data}. Table~\ref{table1} compares key characteristics of the datasets. Figure~\ref{fig:data_description} illustrates each dataset and shape generation heuristic. Note that the purpose of involving the two datasets is to corroborate the versatility of our t-METASET framework, which can accommodate a wide range of datasets born from different methods for different end-uses in a unified way. What we aim to provide is quality assessment of subsets \textit{within} one of the datasets, \textit{not across} them. In addition, while all the datasets in the original references provide the homogenized properties, we assume in all the upcoming numerical experiments that only the shapes are given, \textit{without any property evaluated} a priori. $\mathcal{D}_{TO}$ is publicly available for download at \href{https://ideal.mech.northwestern.edu/research/software/}{https://ideal.mech.northwestern.edu/research/software/}.

\begin{table*}[t]
\caption{Dataset Description}
\begin{center}
\label{table1}
\begin{tabular}{  c  c  c  c   }
\hline
  &  $\mathcal{D}_{lat}$~\cite{wang2022data} & $\mathcal{D}_{mix}$~\cite{chan2022remixing} & $\mathcal{D}_{TO}$~\cite{wang2020data} \\
\hline
Cardinality         &  9,882           & 57,000                                & 88,180 \\
Shape primitive     & Bar               & SDF of basis unit cell                          & N/A (used TO) \\
Shape population      & Parametric sweep                                 & \begin{tabular}[c]{@{}c@{}}Continuous sampling of   \\ basis weights \&  Blending\end{tabular} & \begin{tabular}[c]{@{}c@{}}Stochastic shape perturbation\\ \& Iterative sampling\end{tabular} \\
Topological freedom & Predefined        & Quasi-free                   & Free \\
Property            &
\begin{tabular}[c]{@{}c@{}}
$\{ C_{11}, C_{12}, C_{22},$  \\ $  C_{13}, C_{23}, C_{33} \} $
\end{tabular}       
& $\{ C_{11}, C_{12}, C_{22} \}$ 
& $\{ C_{11}, C_{12}, C_{22} \}$ \\

FEM discretization  & $100\times 100$ & $50\times 50$ & $50\times 50$ \\ 
FEM solver &
\multicolumn{3}{c}{Energy-based homogenization~\cite{xia2015design, andreassen2014determine}} \\

\hline
\end{tabular}
\end{center}
\end{table*}

\subsection{Diversity Metric: Distance Gain}
\label{distance_gain}
We devise an interpretable diversity metric for assessing the capability of t-METASET against benchmark sampling. 
In the literature of DDMD, Chan et al.~\cite{chan2021metaset} compared the determinant of jointly diverse subsets' similarity kernels against those of $iid$ replicates, following the usual practice of reporting set diversity in the DPP literature~\cite{kulesza2012determinantal} as the metric to quantify the efficiency of the proposed downsampling. We point out possible issues of using either similarity or determinant for diversity evaluation: (i) similarity values $s_{ij}$ depend on data preprocessing; (ii) a decreasing transformation from distance to similarity $s_{ij}=T(h(\boldsymbol{x}_i, \boldsymbol{x}_j))$ for constructing DPP kernels also involves arbitrary scaling, depending on the type of associated transformation $T$ and their tuning parameters (e.g., the bandwidth $\sigma_L$ of Gaussian kernels in Eq.~\ref{eq: Gaussian similarity}); (iii) the raw values of both similarity and determinant enable the ``better or worse” type comparison yet lack intuitive interpretation on ``\textit{how much better or worse}”. 

To this end, we propose a distance-based metric that is more interpretable and less arbitrary. Given a dataset $\mathcal{D}$, we compute the mean Euclidean distance $\overline{d}$ of pairwise distances of attributes (shape/property) as
\begin{equation}
\begin{split}
\overline{d}(\mathcal{D})&= \frac{  1 }{|\mathcal{D}|^2}  {   \sum_{j=1}^{|\mathcal{D}|} \sum_{i=1}^{|\mathcal{D}|}  h({\boldsymbol{x}}_i, {\boldsymbol{x}}_j)    }= \frac{  1 }{|\mathcal{D}|^2}  {   \sum_{j=1}^{|\mathcal{D}|} \sum_{i=1}^{|\mathcal{D}|} \sqrt{\| {\boldsymbol{x}}_i-{\boldsymbol{x}}_j \|^2}   }.
\label{eq:mean_euc_dist}
\end{split}
\end{equation}
Intuitively, the larger $\overline{d}(\mathcal{D})$ is, the more diverse $\mathcal{D}$ is. Since this mean metric still depends on data preprocessing, as similarity does, the key idea herein is to normalize $\overline{d}(\mathcal{D})$ with that of an $iid$ counterpart $\overline{d}(\mathcal{D}_{iid})$ with the same cardinality $|\mathcal{D}_{iid}|=|\mathcal{D}|$ so that data preprocessing does not affect it. To account for the stochasticity of $iid$ realizations, we generate $n_{rep}=30$ replicates, take the mean of each mean distance, and compute the relative \textit{gain} $h_G$ as:
\begin{equation}
h_G= \frac{ \overline{d}(\mathcal{D})  }{  \frac{1}{n_{rep}} \sum_{t=1}^{n_{rep}}  \overline{d}((\mathcal{D}_{iid})_l)  },
\label{eq:diversity_metric}
\end{equation}
where $(\mathcal{D}_{iid})_l$ denotes the $l$-th $iid$ replicate with $|(\mathcal{D}_{iid})_l|=|\mathcal{D}|$. We call the metric \textit{distance gain}, as it \textit{relatively} gauges how much more diverse a given set is compared to a set of $iid$ samples. For example, the gain of 1.5 given a property set $\mathcal{P}$ implies that the Euclidean distances between property pairs are 1.5 times larger on average than those of $\mathcal{P}_{iid}$ in the property space. The proposed metric offers an intuitive interpretation based on distance, avoids the dependency on both data scaling and distance-to-similarity transformation, and thus offers a means for consistent diversity evaluation of a given dataset. In addition, the metric generalizes to sequential sampling with $h_G^{(t)}$ at the $t$-th iteration as well, allowing quantitative assessment across datasets at different iterations (i.e., different sizes). Hence, we report all the upcoming results based on the distance gain proposed.

\subsection{Scenario I: Diversity Only}
\label{diversity_only}
\begin{figure*}[t]
\centering
\includegraphics[width=0.8\linewidth]{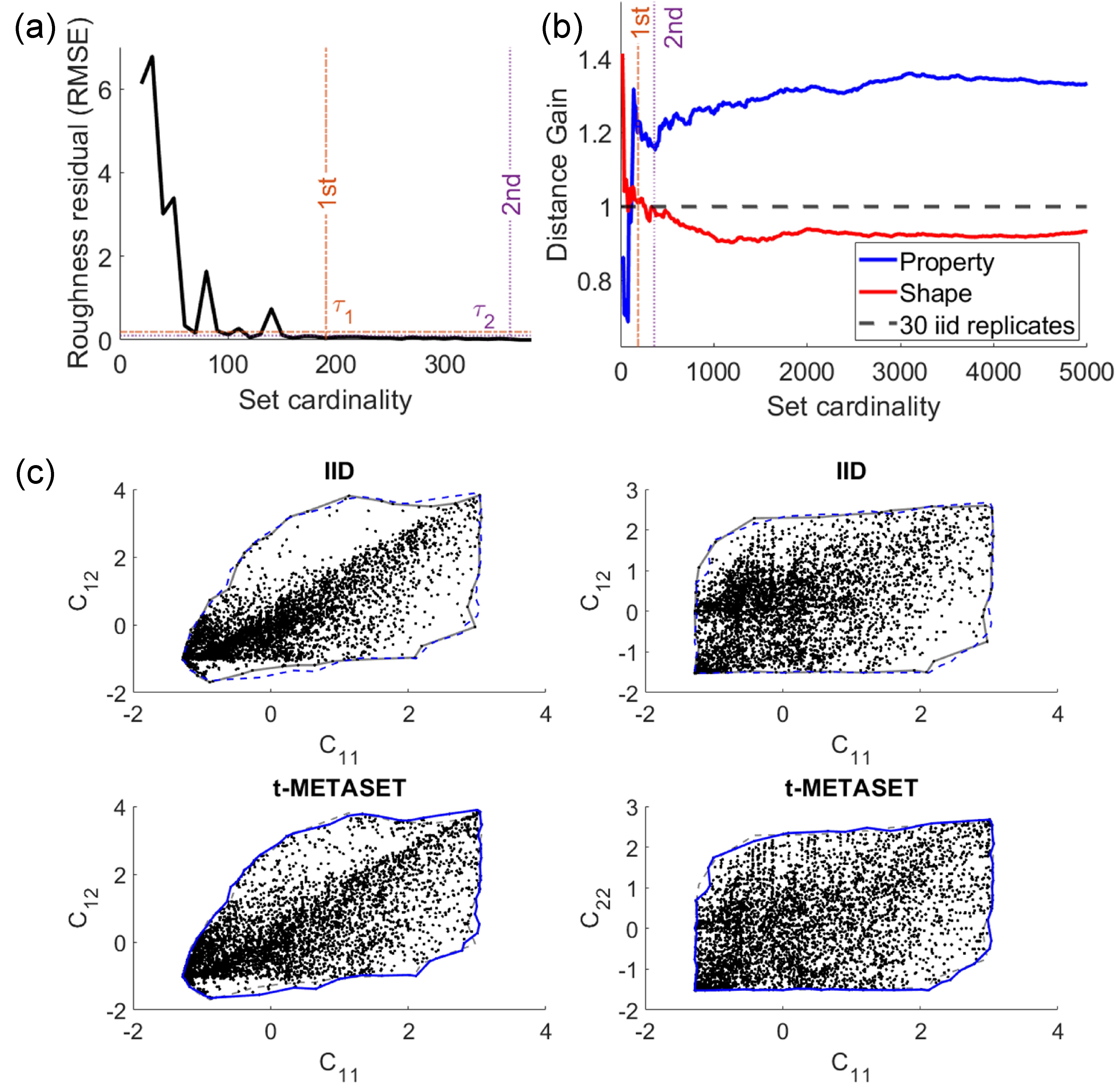}
\caption{Scenario I result for $\mathcal{D}_{TO}$~\cite{wang2020data}. (a) History of roughness residual. The two vertical lines indicate the first and second roughness convergence, respectively. (b) History of distance gains. The horizontal dotted line denotes the distance gain of 30 \textit{iid} replicates, which is unity by the definition in Section~\ref{distance_gain}. The two vertical lines indicate the first ($\tau_1$) and second ($\tau_2$) convergence, respectively. (c) Property distribution in projected property space.}
\label{fig:scenario I}
\end{figure*}

Figure~\ref{fig:scenario I} shows the t-METASET results applied to $D_{TO}$ only based on diversity. From Figure~\ref{fig:scenario I}(a), we observe the evolution of the distance gain as a relative proxy for set diversity at each iteration. At Stage I, the proposed sampling solely relies on shape diversity. The shape gain exceeds unity at the early stage, meaning the exploration by t-METASET shows better shape diversity than that of the $iid$ replicates. Meanwhile, the property diversity of t-METASET is even less than the $iid$ counterpart; this is another evidence that shape diversity barely contributes to property diversity~\cite{chan2021metaset}. During this transient stage, t-METASET keeps monitoring the residual of roughness parameters. Figure~\ref{fig:scenario I}(b) shows the history up to a few hundred observations; the residuals with little data stay unstable, indicating large fluctuations of the hyperparameters. The mild convergence defined by $\tau_1$ occurs at the 19-th iteration with $10\times19=190$ observations. This is approximately twice larger than the rule-of-thumb for the initial space-filling design: $D_\mathbf{z} \times 10=100$~\cite{loeppky2009choosing}. Rigorous comparison between our pairwise initial exploration and space-filling design (e.g., Latin hypercube sampling~\cite{loh1996latin}) is future work.

Once the first convergence criterion on the GP roughness $\boldsymbol{\omega}$ is met, t-METASET starts to respect the GP prediction and, by extension, the RFF of the estimated property DPP kernel as well.
During Stage II, shape diversity decreases to less than unity. This implies that pursuing property diversity compromises shape diversity. After about 300 iterations, each gain seems to stabilize with minute fluctuations, and reach a plateau of about 1.3 for property and 0.95 for shape, respectively. Beyond the maximum iteration set as 500, we forecast that the mean of property Euclidean distances -- the numerator of property gain -- will eventually decrease because: (i) we have finite $|D_{TO}|=88,180$ shapes to sample from; (ii) the property gamut $\partial \Omega^{(t)}_{\boldsymbol{p}}$ at the $t$-th iteration incrementally grows yet ultimately converges to the finite gamut as $\partial \Omega^{(t)}_{\boldsymbol{p}} \rightarrow \partial \Omega^{*}_{\boldsymbol{p}}$, where $\partial \Omega^{*}_{\boldsymbol{p}}$ denotes the property gamut of fully observed $D_{TO}$, which obviously exists yet is unknown in our scenarios; (iii) adding more data points within the confined boundary $ \partial \Omega^{*}_{\boldsymbol{p}}$ would decrease pairwise distances on average. The convergence behavior of the numerator of the property gain may possibly give a hint to answering the fundamental research question in data-driven design: ``\textit{How much data do we need?}”. In addition, adjusting the batch composition -- the ratio of property versus shape -- would lead to different results. The parameter study on $\epsilon$ is addressed and discussed in Section~\ref{Scenario III: Joint Diversity}. 

Figure~\ref{fig:scenario I}(c) shows a qualitative view of the resulting property distributions. Figure~\ref{fig:scenario I}(c) shows the data distribution in the projected property space, whose property components have been standardized. In the $C_{11}$-$C_{12}$ space, the $iid$ realization shows significant bias on the southeast region near $ [-1 \leq C_{11} \leq 1] \times [ -1.5 \leq C_{12} \leq 1 ] $, whereas only tiny samples are located on the upper region. Other 3,000-size $iid$ realizations also result in property bias; local details are different, but the overall trend of distributional bias is more or less the same. On the other hand, the property distribution of t-METASET shows significantly reduced bias in the property spaces, in terms of projected pairwise distances and the property gamut $\partial \Omega_{\boldsymbol{p}}$ as well.

\subsection{Scenario II: Quality-Weighted Diversity (Task-Aware Generation)}
\label{Scenario II: Quality-Weighted Diversity}
Regarding task-aware acquisition of datasets, the scope of this work is dedicated to \textit{pointwise} quality, where the task-related ``value” of each observation is modeled based on a score function. It can be a function of properties (e.g., stiffness anisotropy), shape (e.g., boundary smoothness), or both (stiffness-to-mass ratio). With proper formulation and scaling, the quality function can be included in t-METASET as a secondary sampling criterion. We present two examples, each of which involves either (i) only property (Section~\ref{Scenario II-1}) or (ii) both shape and property (Section~\ref{Scenario II-2}). All the results in this subsection assume the maximum cardinality is fixed as $|\bigcup_{t=1}^{t_{max}} \mathcal{B}^{(t)}|=3,000$.

\subsubsection{Task II-1: Promoting High Stiffness-to-Mass Ratio}
\label{Scenario II-1}
Outstanding stiffness-to-mass ratio is one of the key advantages of mechanical metamaterial systems compared to conventional structures~\cite{yu2018mechanical}. If lightweight design is of interest, users could attempt to prioritize observations with high stiffness-to-mass ratios. We take $C_{11}$ as an example with an associated score $q(\cdot)$ formulated as
\begin{equation}
q_1(\boldsymbol{z},\widehat{{\boldsymbol{p}}})=\frac{\widehat{C_{11}}}{v_f+\delta}
\label{eq:e2vf}
\end{equation}
where $v_f$ is the volume fraction of a given binary shape $I(x, y)$ implicitly associated with $\boldsymbol{z}$, and $\delta$ is a small positive number to avoid singularity. Here, we use raw (not standardized) values of $C_{11}$ to ensure that all the values are nonnegative. Note that the property $\widehat{\boldsymbol{p}}$ takes both (i) ground-truth properties from the finite element analysis and (ii) predicted properties from the regressor $\mathcal{GP}$. To accommodate various datasets at different scales without manual scaling, we standardize $q_1$ into $q_1'$. Then it is passed to the following Sigmoid transformation: 
\begin{equation}
a_1(\cdot)=1-\frac{1}{({1+\text{exp}(-20(\cdot)))}},
\label{eq:Sigmoid1}
\end{equation}
where $a_1(\cdot)$ is the decreasing Sigmoid activation function. To accommodate the design attributes associated with the quality function $a_1(q')$, the RFF $V_{\boldsymbol{p}}$ of the property diversity kernel $\widehat{L_{\boldsymbol{P}}}$ has the pointwise quality on board according to Eq.~\ref{eq:rff_final}.

\begin{figure*}[t]
\centering
\includegraphics[width=0.8\linewidth]{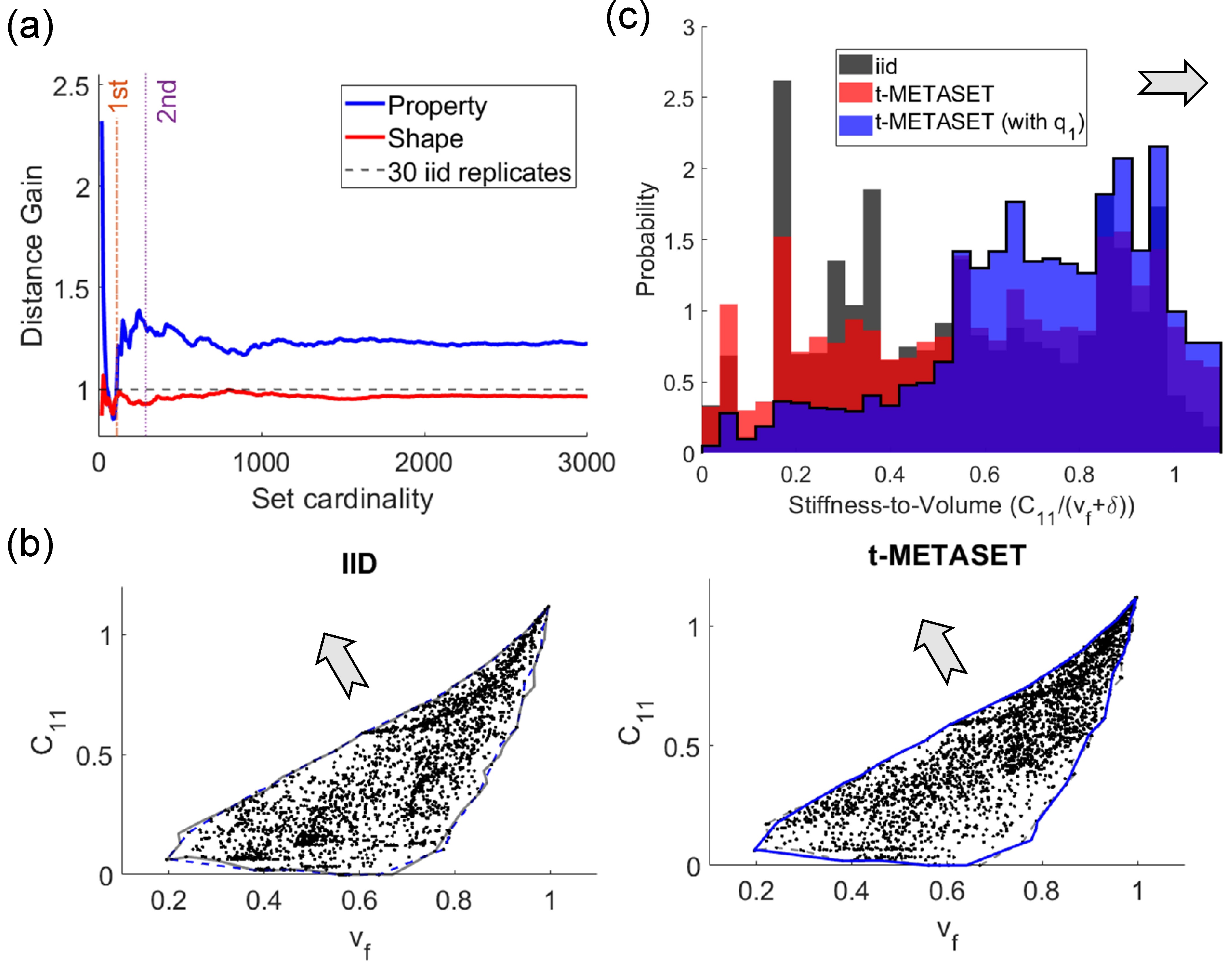}
\caption{Task II-1 (stiffness-to-mass ratio) result for $\mathcal{D}_{mix}$~\cite{chan2022remixing}. Each arrow indicates the distributional bias preferred by the given task. (a) History of distance gains. (b) Data distribution: (left) \textit{iid} and (right) t-METASET with the quality of interest ($q_1$). (c) Histogram of task-related quality (stiffness-to-volume): $iid$ (gray); t-METASET (red); t-METASET with quality $q_1$ (blue).}
\label{fig:scenario II-1}
\end{figure*}

Figure~\ref{fig:scenario II-1} presents the result for $\mathcal{D}_{mix}$. As indicated by the arrow, the quality function aims to bias the distribution in the $C_{11}$-$v_f$ space towards the northwest direction. In Figure~\ref{fig:scenario II-1}(b), the resulting distribution of t-METASET shows an even stronger bias to the upper region than that of the $iid$ replicates, whereas the data points near the bottom right gamut are more sparse. Figure~\ref{fig:scenario II-1}(c) provides even more intuitive evidence: t-METASET without the quality function does not show distributional difference with the $iid$ case. In contrast, the quality-based t-METASET leads to the strongly biased distribution -- virtually opposite to the $iid$ one -- congruent with the enforced quality over high stiffness-to-volume ratio. Both plots corroborate that t-METASET can accommodate the preference of high stiffness-to-volume ratio, \textit{even when starting with no property at all}. Along the way, t-METASET addresses property diversity as well, as indicated by the distance gain of property that exceeds unity (Figure~\ref{fig:scenario II-1}(a)).

\subsubsection{Task II-2: Promoting High Stiffness Anisotropy}
\label{Scenario II-2}
Property anisotropy of unit cells is another key quality that mechanical metamaterials could leverage to achieve strong directional performances at system levels. With $\mathcal{S}_{TO}$, we attempt to deliberately bias the property distribution towards strong elastic anisotropy between $C_{11}$ and $C_{22}$. We devise the anisotropy index as an associated quality function:
\begin{equation}
q_2(\boldsymbol{z},\widehat{{\boldsymbol{p}}})=\frac{|\arctan(\widehat{C_{22}}/\widehat{C_{11}})-\pi/4|}{\pi/4}, 
\label{eq:aniso}
\end{equation}
where $\widehat{C_{11}}$ and $\widehat{C_{22}}$ denote the raw non-negative elastic constants predicted by the GP model, without any normalization; $\arctan(\widehat{C_{22}}/\widehat{C_{11}}) \in [0, \pi/2]$ is the polar angle in the $C_{11}$-$C_{22}$ space; if isotropic (i.e.,  $C_{11}=C_{22}$), the index is 0, whereas either $C_{22}/C_{11} \rightarrow 0^+$ or $C_{22}/C_{11} \rightarrow \infty$, the index goes to 1. By the definition, the quality function ranges within $[0, 1]$. Without further scaling, we directly pass it to a monotonically increasing Sigmoid activation: 
\begin{equation}
a_2(\cdot)=\frac{1}{1+\text{exp}(-20((\cdot)-0.5))}.
\label{eq:Sigmoid2}
\end{equation}
Similar to the first example above, $a_2(q_2)$ is incorporated into the RFF of property through Eq.~\ref{eq:rff_final}.

\begin{figure*}[t]
\centering
\includegraphics[width=0.8\linewidth]{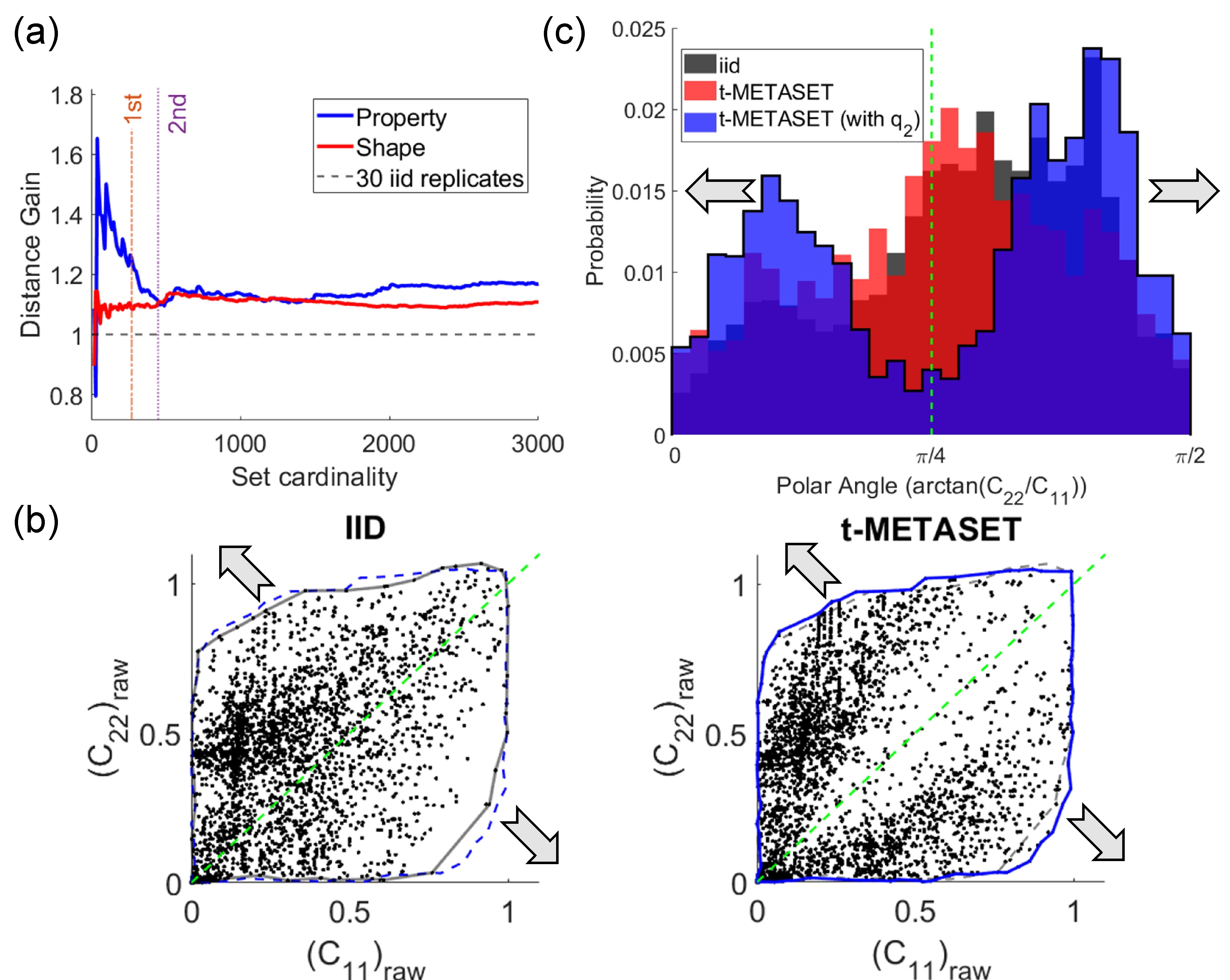}
\caption{Task II-2 (stiffness anisotropy) result for $\mathcal{D}_{TO}$~\cite{wang2020data}. Each arrow indicates the distributional bias preferred by the given task. The green dotted line denotes the subregion where $C_{22}=C_{11}$, least preferred by the given task. (a) History of distance gains. (b) Data distribution: (left) \textit{iid} and (right) t-METASET with the quality of interest ($q_2$). (c) Histogram of task-related quality (polar angle): $iid$ (gray); t-METASET (red); t-METASET with quality $q_2$ (blue).}
\label{fig:scenario II-2}
\end{figure*}

Figure~\ref{fig:scenario II-2} illustrates the result for $\mathcal{D}_{TO}$ under the anisotropy preference. The two arrows indicate the bias direction of interest: Samples with isotropic elasticity on the line $C_{22}=C_{11}$, denoted as the green dotted line, are least preferred. From the scatter plot of Figure~\ref{fig:scenario II-2}(a), the distribution of t-METASET exhibits clear bias towards the preferred direction compared to the $iid$ case, while samples near the isotropic line are sparse except near the origin. The trend is even more apparent in the histograms of Figure~\ref{fig:scenario II-2}(b): both the results from $iid$ and vanilla t-METASET share a similar distribution in terms of polar angle. In contrast, task-aware t-METASET exhibits a bimodal distribution that is highly skewed to either $0$ or $\pi/2$.

In Figure~\ref{fig:scenario II-2}(a), we recognize an interesting point that reveals the power of t-METASET: unlike the other cases introduced, the shape gain also exceeds unity at the plateau stage, at mild cost of the property gain. 
Note that we did \textit{not} enforce the framework to assign more resources on shape diversity. The quality function $q_2(\cdot, \cdot)$ has been defined over only the two properties $C_{11}$ and $C_{22}$, \textit{not} shape. Furthermore, during Stage II, t-METASET can take only two samples from shape diversity in each batch due to the setting $\epsilon=0.8$, commonly shared by the other cases introduced. This indicates that: the decent exploration in the shape space -- the shape gain comparable to the property gain during Stage II -- is \textit{what t-METASET autonomously decided via active learning to fulfill the mission specified by the given task}. The result demonstrates the ability of t-METASET to, given a large-scale dataset and on-demand design quality, decide how to properly tradeoff distributional biases in shape/property space, thereby efficiently addressing the design goals without human supervision.


We emphasize that the two results came from the same algorithmic settings of t-METASET shared with the other cases, except for the quality functions. Hence, the two case studies, investigated with respect to different datasets and different quality functions, demonstrate that t-METASET has fulfilled the mission: growing task-aware yet balanced datasets by active learning.

\subsection{Scenario III: Joint Diversity}
\label{Scenario III: Joint Diversity}
The proposed t-METASET can tune shape-property joint diversity when building datasets. Chan et al.~\cite{chan2021metaset} demonstrated that, given a \textit{fully observed} dataset, the DPP-based sampling method can identify representative subsets with adjustable joint diversity~\cite{chan2021metaset}. It is grounded on the fact that any linear combination of PSD shape and property kernels can create a joint diversity kernel $L_J=(1-\epsilon)L_{\boldsymbol{s}}+\epsilon L_{\boldsymbol{p}}$ that is also PSD, where $L_{\boldsymbol{s}}$ is a shape similarity kernel involving a shape descriptor $\boldsymbol{s}$. Yet the linear combination approach does not apply to our proposed t-METASET, driven by the RFF $V$, because the linear combination of the feature $V$ does not guarantee the resulting joint kernel to be PSD.

\begin{figure*}[t]
\centering
\includegraphics[width=0.8\linewidth]{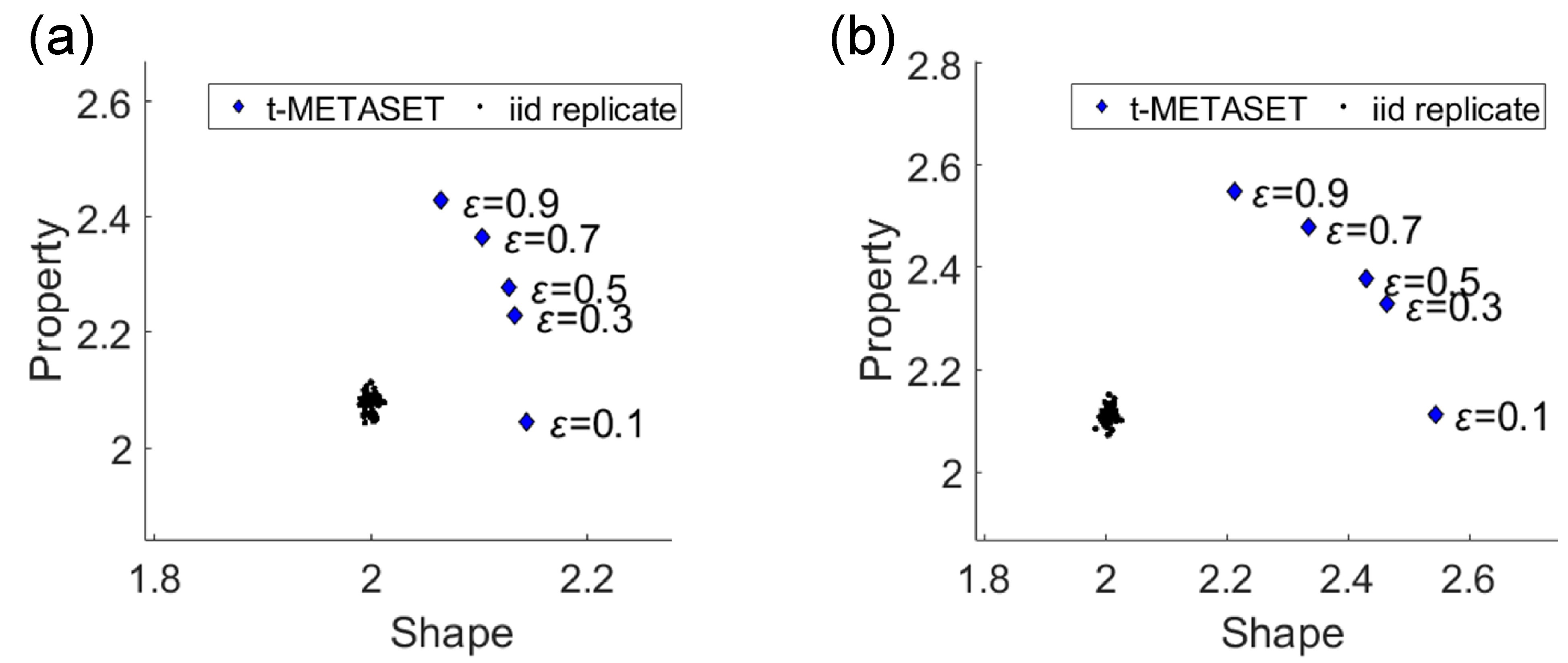}
\caption{Scenario III (joint diversity) results for $\mathcal{D}_{mix}$~\cite{chan2022remixing} and $\mathcal{D}_{TO}$~\cite{wang2020data}. $\epsilon$ denotes the ratio of property samples in each batch. (a) Mean Euclidean distances for $\mathcal{D}_{mix}$. (b) Mean Euclidean distances for $\mathcal{D}_{TO}$.}
\label{fig:scenario III}
\end{figure*}

Instead, our framework tunes joint diversity by adjusting the shape/property sampling ratio $\epsilon$ of a batch. Figure~\ref{fig:scenario III} shows the parameter study over the batch composition $\epsilon$ with respect to $D_{mix}$ and $D_{TO}$ with $|\bigcup_{t=1}^{t_{max}} \mathcal{B}^{(t)}|=5,000$. Both results manifest (i) better average diversity in terms of Euclidean distances than that of the $iid$ replicates, and (ii) the tradeoff between shape diversity and property diversity. Additionally, the results support the previous finding that the correlation between shape diversity and property diversity is near-zero~\cite{chan2021metaset}. The substantial distinction of t-METASET lies in: We \textit{sequentially} achieve the jointly diverse datasets, \textit{beginning from scratch in terms of property}. In addition, t-METASET allows users to dynamically adjust $\epsilon$ as well, based on either real-time monitoring over diversity gains or user-defined criteria. This capacity could possibly help designers steer the sequential data acquisition at will, especially if growing a large-scale dataset ($\sim O(10^4)$) is of interest, since applying a single sampling criterion over the whole generation procedure might not necessarily result in the best dataset for given design tasks.

\subsection{Algorithm Efficiency}
\begin{figure}[t]
\centering
\includegraphics[width=0.7\linewidth]{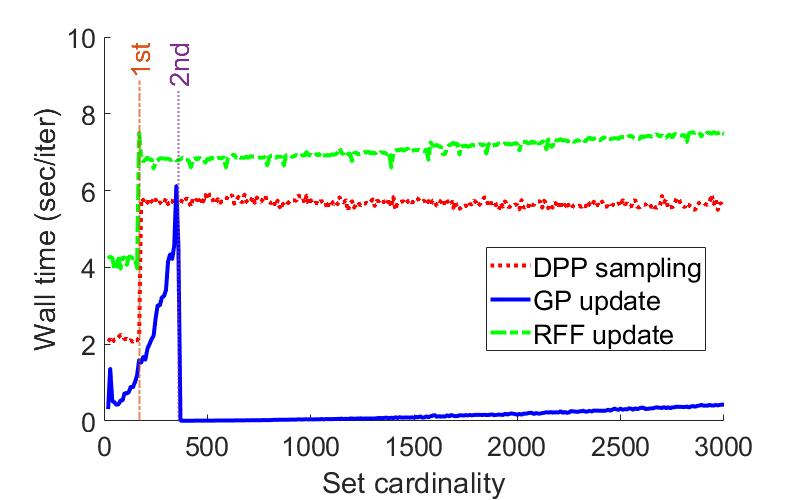}
\caption{Elapsed real time measurement of a replicate of $\mathcal{D}_{TO}$ as a function of dataset size. ``1st” and ``2nd” denote the sampling transition points corresponding to the thresholds $\tau_1$ and $\tau_2$, respectively.}
\label{fig:computation time}
\end{figure}

The t-METASET procedure is, in essence, a sequential decision making problem that selects from a given large pool of instances $(\sim \mathcal{O}(10^4))$. Its scalability comes primarily from the RFF based kernel approximation, and secondarily from the compact 10-D shape descriptor distilled from the VAE training. To give readers a glimpse of the scalability of the proposed data acquisition, Fig.~\ref{fig:computation time} shows the history of wall time (i.e., elapsed real time) per iteration for $\mathcal{D}_{TO}$, where approximately 88k instances are included. The test was run using a desktop with Intel(R) Xeon(R) W-2295 CPU @ 3.00GHz, 18 cores/36 threads, RAM 256 Gb.

For each iteration, we look into the trend of the wall time based on three key steps: (i) GP updating, (ii) DPP sampling, and (iii) RFF updating. In the early stages, the incurred time for the GP update escalates rapidly over the dataset size, dominated by the inversion of covariance matrices. Once the second condition of roughness convergence is met ($\Delta^{(t)} \leq \tau_2$; denoted as ``2nd” in Fig.~\ref{fig:computation time}), the improvements of GP updates over new batches become marginal. The sequential updates are then replaced by the preposterior analysis~\cite{van2021scalable}, whose computational cost gradually increases as the cardinality grows. Meanwhile, the conditional \textit{k}-DPP for sequential diversity sampling takes up a moderate portion of time at each iteration. At the start-up phase, the DPP sampling is performed only once per iteration for the sampling based on shape diversity. As of Stage II, DPP sampling runs twice; once based on property diversity (which is optionally weighted by quality), followed by the other based on shape diversity. The incurred wall time shows little dependence on cardinality $|\mathcal{D}^{(t)}|$, as its time complexity primarily depends on the number of replicates in RFF (i.e., $D_V$). Lastly, the main computational overhead of the t-METASET procedure involves updating the RFF. In Stage I, only the shape RFF is updated. Once the first convergence of roughness parameters is met ($\Delta^{(t)} \leq \tau_1$; denoted as ``1st” in Fig.~\ref{fig:computation time}), the property RFF is calculated and updated. Thereafter, every iteration involves (i) updating the shape RFF, (ii) constructing a property RFF that mirrors the latest GP, and (iii) conditioning the property RFF on the sample collected up to the current iteration.

\section{Conclusion}

We presented the task-aware METASET (t-METASET) framework dedicated to metamaterials data acquisition congruent with user-defined design tasks. Distinctly, t-METASET specializes in a data-driven scenario that designers often encounter in early stages of DDMD: a massive shape library has been prepared with no properties observed for a new design case. The central idea of t-METASET for building a task-aware dataset, in general, is to (i) leverage a compact yet 
expressive shape descriptor (e.g., VAE latent representation) for shape dimension reduction, (ii) sequentially update a sparse regressor (e.g., GP) for nonlinear regression with sparse observations, and (iii) sequentially sample in the shape descriptor space based on estimated property diversity and estimated quality (e.g., DPP) for distribution control over shape and property. t-METASET contributes to the design field by: (i) proposing a data acquisition method \textit{at early data-driven stages under large epistemic uncertainty}, (ii) \textit{sequentially} combating \textit{property bias}, and (iii) accommodating \textit{task-aware design quality} as well. Starting without evaluated properties, all the results tested on two large-scale metamaterial datasets ($\mathcal{D}_{mix}$ and $\mathcal{D}_{TO}$) were automatically achieved by t-METASET in three different scenarios without human supervision. We argue t-METASET can handle a variety of image-based datasets for design in general, by virtue of scalability, modularity, task-aware data customizability, and independence from both shape generation heuristics and domain knowledge.

Whereas the present scope of t-METASET is dedicated to metamaterials, the framework is applicable to other material systems where the structures, such as microstructure morphology, can be quantified. Three example scenarios in which t-METASET has potential are provided here:
\begin{itemize}[label=\textbullet]
    \item A low-dimensional representation is prescribed by a designer. This applies not only to metamaterials with an explicit parameterization (e.g., the lattice-type building block specified by four parameters~\cite{wang2022data} in \textcolor{blue}{Section 2}), but to other systems as well (e.g., quasi random organic photovoltaic cells represented with a 2-D spectral density function~\cite{farooq2018spectral, iyer2020designing})
    \item A mixed categorial and quantitative representation is given. A key modification in t-METASET would be to replace the vanilla GP with a latent variable Gaussian Process (LVGP)~\cite{zhang2020latent}. An example is the multi-class lattice metamaterial dataset in Ref.~\cite{wang2021data}. Therein, any instance of a material is specified as $\boldsymbol{z}=(c_i, \rho) (i=1, 2, \cdots)$, where a qualitative variable $c_i$ is the class index of lattice-type building blocks, and a quantitative variable $\rho$ is the volume fraction.
    \item No representation is given (the scenario of primary interest in this work). Unsupervised representation learning can be harnessed, as has been employed in this work, to prepare a compact yet expressive descriptor in light of a dataset.
\end{itemize}
Thus we argue our framework can address other systems, such as those represented by user-defined descriptors or by pixels/voxels, \textit{beyond} metamaterials. A possible issue, in particular when dealing with a system with 3-D volume elements (e.g., polymer nanocomposite), is that the dimensionality of a shape descriptor could be too large for a vanilla GP to handle, even after dimension reduction. Two workarounds for this case are: (i) employing extended GPs dedicated to high-dimensional data~\cite{rana2017high, tripathy2016gaussian} or (ii) using other surrogates with more modeling capability (e.g., a moderately sized neural network). 

The imperative future work is inference-level validation of dataset quality, which aims to shed light on the downstream impact of data quality at the deployment stage of data-driven models. Among a plethora of such models, we are particularly interested in conditional generative models~\cite{mirza2014conditional, sohn2015learning} due to their on-the-fly inverse design capability, which is expected to be highly sensitive to data quality~\cite{zheng2021continuous, heyrani2021pcdgan}. The validation would further demonstrate the efficacy of t-METASET at the downstream stages of DDMD, in addition to at the intuitive metric level we have shown. Moreover, we point out two interesting topics to be explored: (i) the proposed diversity gain as a termination indicator of data generation, which could offer insight into ``\textit{how much data?}” (detailed in Section~\ref{diversity_only}), and (ii) quantitative comparison between the quality-weighted diversity sampling (Section~\ref{Quality-Weighted Diversity}) presented in this work and BO~\cite{snoek2012practical,tao2021multi}.

Through producing and sharing open-source datasets, t-METASET ultimately aims to (i) provide a methodological guideline on how to generate a dataset that can meet individual needs, (ii) publicly offer datasets as a reference to a variety of benchmark design problems in different domains, and (iii) help designers diagnose their dataset quality on their own. This lays a solid foundation for the future advancement of data-driven design.

\section*{Acknowledgment}
We acknowledge funding support from the National Science Foundation (NSF) through the CSSI program (Award \text{\#} OAC 1835782).



\bibliographystyle{unsrt}  
\bibliography{main}  


\end{document}